\documentclass[aps, prd, amsmath, floats, floatfix, twocolumn, superscriptaddress, nofootinbib, showpacs]{revtex4}

\usepackage{graphicx,color}
\usepackage{latexsym}
\usepackage{amsmath,amssymb}        
\usepackage[draft=false]{hyperref}
\usepackage{mathrsfs}
\usepackage{comment}
\usepackage{soul}
\usepackage{mathrsfs}
\definecolor{orange}{rgb}{1,0.5,0}
\DeclareSymbolFontAlphabet{\mathrsfs}{rsfs}
\DeclareMathAlphabet{\mathcal}{OMS}{cmsy}{m}{n}

\usepackage{ulem}

\begin{document}
\title{Short Review of the main achievements of the\\ Scalar Field, Fuzzy, Ultralight, Wave, BEC Dark Matter model}

\author{Tonatiuh Matos}\email{tonatiuh.matos@cinvestav.mx}
\affiliation{Departamento de F\'isica, Centro de Investigaci\'on y de Estudios Avanzados del IPN, A.P. 14-740, 07000 M\'exico D.F.,
  M\'exico.}

  \author{Luis A. Ure\~na-L\'opez}
\email{lurena@ugto.mx}
\affiliation{Departamento de F\'isica, DCI, Campus Le\'on, Universidad de Guanajuato, 37150, Le\'on, Guanajuato, M\'exico.}

\author{Jae-Weon Lee}
\email{scikid@jwu.ac.kr}
\affiliation{Department of Electrical and Electronic Engineering, Jungwon University, 85 Munmu-ro, Goesan-eup, Goesan-gun, Chungcheongbuk-do, 28024, Korea.}

\begin{abstract}
The Scalar Field Dark Matter model has been known in various ways throughout its history; Fuzzy, BEC, Wave, Ultralight, Axion-like Dark Matter, etc. All of them consist in proposing that the dark matter of the universe is a spinless field $\Phi$ that follows the Klein-Gordon (KG) equation of motion $\Box\Phi-dV/d\Phi=0$, for a given scalar field potential $V$. The difference between different models is sometimes the choice of the scalar field potential $V$. In the literature we find that people usually work in the nonrelativistic, weak-field limit of the KG equation where it transforms into the Schr\"odinger equation and the Einstein equations into the Poisson equation, reducing the KG-Einstein system, to the Schr\"odinger-Poisson system. In this paper, we review some of the most interesting achievements of this model from the historical point of view and its comparison with observations, showing that this model could be the last answer to the question about the nature of dark matter in the universe.  
\end{abstract}

\maketitle
\section{Introduction}

  The Scalar Field Dark Matter (SFDM) model proposes that the dark matter of the universe is a spinless particle $\Phi$ fulfilling the Klein-Gordon equation 
  \begin{equation}
      \Box\Phi-\frac{dV}{d\Phi}=0
  \end{equation}
  where $\Box$ is the D'Alambertian operator in a curved spacetime, whose source of Einstein's equations is the Scalar Field (SF), where the SF can be neutral (real) or charged (complex). Therefore, strictly speaking, we have to solve the Einstein-KG system. However, in galaxies and in general in the universe, systems are non-relativistic, and gravitation is weak enough so that the Einstein-KG system can be reduced to the Schr\"odinger-Poisson system, which in general is much easier to solve.

  The origin of SF is unknown since we do not have a definitive unifying theory of all interactions. The easiest way to understand its origin is to add an SF term to the Standard Model (SM) of particles as we do with the rest of the terms of this model. There are other proposals, such as that the SF comes from superstring theory or that it has a QCD-type origin but with ultralight mass, etc. We will treat it here in a generic way, and we will focus only on the results as a dark matter candidate, leaving its origin for future work.

  Dark matter behaves in a way similar to dust, as in the case of Cold Dark Matter (CDM), and SF mimics dust when it oscillates at the minimum of a potential~\cite{Turner:1983he}. Therefore, the most popular SF potentials are those in which the SF potential $V$ has a minimum. It is convenient to expand the SF potential into an even series of $\Phi$ as
  \begin{equation}
      V=V_0+\frac{1}{2}m^2\Phi^2+\frac{\lambda}{4}\Phi^4+\cdots
  \end{equation}
   where we can interpret $V_0$ as the expectation value of the SF vacuum, $m$ is the mass and $\lambda$ the self-interaction parameter, where for complex SF the $\Phi^2$ changes by $|\Phi|^2$. This expansion is generic as long as we do not know the correct form of $V$. However, there are some proposals for $V$ coming from some theories such as QCD axions, with the SF potential $V=f\sin(b\Phi)$ or $V =A(\cosh(b\Phi)- 1)$ \cite{Matos:2000ss}, \cite{Sahni:1999qe}, derived from the superstring theory.

   To the best of our knowledge, the idea that an SF could be dark matter of the universe began in 1983 by Baldeschi et al.~\cite{Baldeschi:1983mq}, where they fit the rotation curves of galaxies using boson and fermion particles. In 1990 the authors in~\cite{Press:1989id} presented a model in which a quartic potential $V=V_0(1-\lambda|\Phi|^2)^2$ can explain the large-scale structure of the universe; this SF covers the missing mass of the universe. The first comparison of the model with real galaxies was made in 1992~\cite{Sin:1992bg,Ji:1994xh}, where the authors proposed a Bose gas to fit the rotation curves of galaxies, with a mass of the SFDM of the order of $10^{-23}$eV. The first numerical simulations with SF to form galaxies were presented in 1993 in~\cite{Widrow:1993qq}, where they obtained the shape of a galaxy using the Sch\"odiger-Poisson system. 
   
In 1995 \cite{Lee:1995af} suggests that repulsive quartic self-interaction increases the length scale to $O(\sqrt{\lambda} m_P/m^2)$
even for a tiny $\lambda$, where $m_P$ is the Planck mass. An approximate analytic solution for the ground state was obtained
for the Thomas-Fermi limit. Based on the theory of boson stars, the maximum stable central density and the maximum halo mass for the $\lambda=0$ case yielded a bound of $10^{-28} \leq  m \leq 10^{-22} eV$.

   Independently, in 1998, it was proposed in~\cite{Matos:1998vk} that SF can solve the problem of dark matter and that this field can explain the rotation curves of galaxies, initiating the first systematic study of this paradigm.
   Subsequently, the idea of SFDM has been rediscovered many times, such as Fuzzy DM~\cite{Hu:2000ke}, Quintessential DM~\cite{Arbey:2001qi,Arbey:2001jj}; Ultralight DM~\cite{Amendola:2005ad,Lundgren:2010sp}; Bose-Einstein condensate DM~\cite{Boehmer:2007um,Rindler-Daller:2009qyu,Chavanis:2011gm}; Wave DM (\cite{Bray:2010fc,Schive:2014dra}), etc. In 2017 this idea became fashionable and one of the favorite candidates to explain DM after the publication of the article~\cite{Hui:2016ltb}.

   The SFDM model has two parameters, the mass $m=\hat{m}c/\hbar$, where $\hat{m}$ is the mass in grams, and the self-interaction parameter $\lambda$. With these two parameters, it is possible to fit a large number of DM observations into the universe. The objective of this work is to historically list some of its most important achievements, and we will focus on the following. 

\begin{itemize}
    \item Good agreement of the rotation curves of stars and dust around galaxies.
    \item Good agreement with cosmological constrains and Big Bang Nucleosynthesis
    \item An alternative solution to the cusp-core problem.
    \item An alternative solution to the problem of satellites.
    \item The central black holes in galaxies.
    \item This is in excellent agreement with the CMB and MPS cosmological observations.
    \item The agreement with the cosmological numerical simulation.
    \item A natural explanation of the anomalous trajectories of satellites around galaxies.

    Some predictions
    \item The existence of a soliton at the center of galaxies.
    \item The structure formation of galaxies according to cosmological numerical simulations
    \item An alternative explanation to the Fermi Bubbles
    
\end{itemize}

 In what follows, we want to touch on each of these issues by giving a historical development of the paradigm showing how the SFDM model solves or reduces the problem in each of these issues.

\section{Rotation curves}

Completely independently, in 1998 it was proposed by Tonatiuh Matos and Francisco S. Guzm\'an the idea that dark matter (DM) could be a scalar field as a doctoral topic. In this work (\textit{Scalar fields as dark matter in spiral galaxies}, Matos and Guzman, arXiv:gr-qc/9810028~\cite{Matos:1998vk}), was shown that this hypothesis could explain the observed rotation curves of stars and gas around galaxies. As in~\cite{Sin:1992bg}, the authors can fit well the rotation curves of specific galaxies.

However, in 2002 it was noted that there were two problems with the model that had to be dealt with. The first is that galaxies were found to be unstable according to this model. The second is that supermassive black holes in galaxies could swallow the SFDM halo. The second problem will be
described in Sec.~\ref{ch:SMBH}. Next, we describe the first problem.

In \cite{Guzman:2003kt,Guzman:2004wj,Guzman:2006yc} the authors found that the numerical simulations of the collapse of the SFDM do not stabilize, and the collapse continues until it forms a too compact object; the authors call it gravitational cooling. This collapse depends on the mass of the object; for large objects, this gravitational cooling is long enough to explain the existence of galaxy clusters, but for smaller masses it is too short to explain the shape of galaxies.
There are at least two ways out of this problem. The first is to take into account the rotation of the SFDM. If the SFDM has a spin, the spin prevents the galaxy from collapsing. The second way out is to take into account the quantum character of the SFDM and invoke the excited states of the system.

It is possible to derive a Tully-Fisher relation from SFDM as in MOND~\cite{Bray:2014dca,Lee:2019ums}.

Another line of research is that when the SFDM is in a thermal bath and its finite temperature is taken into account, the Schrödinger equation predicts that the SF can have fundamental and excited states. The first attempts at this line of research are~\cite{Matos:2011kn}, using a one-loop SF potential, and~\cite{Harko:2011dz} using the SF expansion. More approaches to this are given in~\cite{Matos:2011pd,Robles:2013dfa}, where the rotation curves are fitted using the ground and excited states of the SFDM. But the most interesting success of this line of research is the fact that the excited states are figure 8-shaped and are capable of explaining the Vast Polar Orbits (VPO), which we will see in a later section.

\section{The Cosmological Constrains}

Since the model fits the rotation curves of galaxies well, it must also be checked whether this model complies with cosmological observations, for example, Big Bang Nucleosynthesis (BBN), Mass and Power Spectrum (MPS), Cosmic Microwave Background (CMB), etc. This was done for the first time in \cite{Matos:2000ss}, where the authors used a modified code CMBfast to obtain the MPS and CMB of the model, and they found more or less agreement with the data found up to this point. At that time, the data were not that good, but it was enough to say that the agreement between the data and the model is good enough for the model to be viable. 
This analysis was then performed several times as the data were improved, showing that the model agrees excellently with the cosmological data; see, for example~\cite{Harko:2011jy,Urena-Lopez:2015gur,Hlozek:2014lca,Cedeno:2017sou}. 

The next observation to be taken into account is the BBN. In~\cite{Li:2013nal} it has been shown that to comply with the BBN constraint, it is necessary for the SF to have a self-interaction $\lambda\Phi^4$, with $\lambda\sim 10^{-90}$ ultra small but not zero. This result is very important because it implies that the SF must have a small self-interaction to satisfy all cosmological constraints, although this very small self-interaction seems to have no observational repercussions at the galactic level.

Furthermore, gravitational waves (GW) impose some restrictions on the SFDM model. In~\cite{Li:2016mmc} it is shown that the GWs produced during inflation can interact with the SFDM and produce a signal detectable by LIGO/Virgo and LISA. This possible interaction imposes some limitations on the reheating temperature of the universe, excluding a range of reheating-temperature parameters for a certain range of mass and self-interaction parameters.

\section{The cusp-core problem}
This problem has been observed since the 1990s, where people realize that CDM numerical simulations predict a huge concentration of DM at the center of galaxies and observations in dwarf galaxies seem to indicate that there is not as much concentration there, that is, the density profile of the DM in the center appears to be constant. There is a lot of discussion on this topic, but observations at the center of galaxies are not compatible with CDM predictions. To solve this problem, CDM needs more physical assumptions to flatten the density profile. There is a huge literature on this. Furthermore, in the satellite galaxies of our neighborhood, the center of these galaxies shows that they all have the same mass of $10^7M_\odot$ within the first $300$pc, for at least six orders of magnitude in luminosity of the satellite galaxies \cite{Strigari:2008ib}. Although it has been confronted with LCDM, its solution is not satisfactory.

The first paper proposing that SFDM could provide a natural solution to this problem was in \cite{Hu:2000ke}, where the authors solve the SFDM equations in one dimension to show that SFDM does not have such concentrations. The idea they put forward is that the SFDM model, which they called Fuzzy Dark Matter, has a quantum character and it is the uncertainty principle that prevents the DM from concentrating on a point, because in that case we are able to locate it. Due to the uncertainty principle, the location of the DM causes the angular momentum to grow and this prevents the concentration of matter in the center. 
The first analytical result on this topic was published in~\cite{Argelia08}, where the authors use approximations to find the density profile of dwarf galaxies and find a fit between observations of real dwarf galaxies and the central density profile predicted by SFDM, showing that this problem is solved naturally for the SFDM model without further assumptions. See also~\cite{Su:2010bj,Harko:2011xw} for an alternative approach.
This result was corroborated in \cite{Schive:2014dra} using 3D numerical simulations, and the authors show that SFDM does indeed have a core density profile as previously predicted by \cite{Hu:2000ke} and \cite{Argelia08} and opened an important line of research for 3D numerical simulations followed by \cite{Du:2016zcv}, \cite{Mocz:2019pyf}, among others. Numerical simulations of SFDM today have the same level of accuracy as CDM, which shows that SFDM is a viable model for DM of the universe.

The observation that all satellite galaxies have the same amount of DM inside of the first 300pc has also been faced using SFDM in~\cite{Lee:2008jp} using numerical simulations, here the authors find that visible matter density of a dwarf galaxy has a universal core size $r_c= 4000$, corresponding to the physical size $r_c \sim 300$pc and the total mass within this size $M_{tot} \sim 0.00019$, corresponding to $4.75 \times 10^7M_\odot$. However, the SFDM is a field and satisfies the Schr\"odinger-Poisson field equations that contain scale invariance. In~\cite{Urena-Lopez:2017tob} using analytical methods, it was shown that this scale invariance implies that all galaxies must have the same mass within $\sim 350$pc, giving a completely natural explanation to the observations of~\cite{Strigari:2008ib}.

\section{The satellites problem}
In the year 2000, Tonatiuh Matos and Luis A. Ureña-López, studied for the first time the SFDM hypothesis from a cosmological point of view. In~\cite{Matos:2000ss} the results were spectacular, finding for the first time that all cosmological observations up to that time were explained within the error bars by SFDM. They showed that the CMB and MPS were in agreement with those observed at the time and began the systematic cosmological study of this paradigm. The main result in~\cite{Matos:2000ss} was that, using the mass of the scalar field as a free parameter, it was shown that the SF has a natural cutoff of the mass power spectrum, which implies that the theoretical number of satellite galaxies is of the order of magnitude of those observed, provided that the mass of the SF is $10^{-22}$eV, coinciding with the mass necessary to explain the rotation curves of galaxies. This result was corroborated many years later by numerical simulations~\cite{Schive:2014dra} and semi-analytical analysis~\cite{Bozek:2014uqa}.

 Simply, the main idea was that perturbations of the scalar field $\delta\Phi$ from the decomposition $\Phi=\Phi_0+\delta\Phi$ form the large-scale fabric of space-time. Here $\Phi_0$ is the SF for the background and depends only on $t$. The point is that this perturbation $\delta\Phi$ follows a damped harmonic equation driven by a force that oscillates with a frequency very similar to the mass of the SF. The damped term essentially depends on the mass of the SF, the scale factor $a$ that determines the redshift of the perturbation, and the wave number of the Fourier transform of the perturbation, that is, the size of the perturbation. When the damping term is in resonance with the oscillating force term, we will have an increasing perturbation. But when they are not in resonance, the damping term will cause the perturbation to decrease and disappear. With this a relationship was found between the size and redshift of the perturbation and the mass of the SF versus the frequency of the driving force. The free parameter $m$ was set using the hitherto known number of galaxies in our neighborhood given the magic number of $m\sim 10^{-22}$eV, according to the same mass value found to fit rotation curves in galaxies.  In fact, the disturbance equation was solved with numerical methods, giving a relationship between all the quantities involved.

 The first work that proved that this result is right was in~\cite{Guzman:2004wj}, where the author found that the gravitational collapse of a SF contains, in fact, this cut-off point of the mass power spectrum. Further in~\cite{Schive:2014dra}, using numerical 3D simulations, the author finally showed that indeed the number of satellite galaxies was on the order of the observed one in our galaxy. See also~\cite{Du:2016zcv,Mocz:2019pyf}.
 
 It is also suggested that the SFDM model exhibits a characteristic size and mass scale consistent with dwarf galaxies and can account for the observed size evolution of very massive compact galaxies in the early universe
 \cite{Lee:2008ux,Lee:2015cos}.
 
\section{The central Black Holes}\label{ch:SMBH}

Galaxies contain a Supermassive Black Hole (SMBH) at their center. However, it is difficult to understand how some SMBHs formed, especially those at high redshifts. If the growth of these SMBH were by accretion, we have to explain how this accretion of matter grows the SMBH from $10^2M_\odot$ to, say, $10^{9}M_\odot$ with an accretion of one $M_\odot $ per year, at high redshifts. There are a few proposals for this, ranging from galaxy collisions to primordial formation of SMBH during the Big Bang. All of them contain pros and cons, and the SFDM offers an alternative explanation of this problem that seems very natural but so far incomplete. The proposal is based on numerical simulations carried out in~\cite{Seidel:1991zh}, where the authors found that the critical mass of collapse of a real scalar field is $M_{crit}=0.6m_{pl}^2/m$, where $m_{pl}$ is the Planck mass and $m$ is the SFDM mass. If we plug $m\sim 10^{-22}$eV into this formula, we find $M_{crit}\sim10^{13}M_\odot$, which is very large. For a complex scalar field, the result is similar \cite{Balakrishna:1997ej}, the authors find that the critical mass of collapse for a complex SF is $M_{crit}=0.1m_{pl}^2/m$. This gives two results; the first is that SFDM galaxy halos have a given natural mass limit, so that beyond this critical mass the SF collapses to form BH. However, at the same time, this large collapsing mass could explain the formation of SMBH at the center of galaxies, provided that a mechanism can be provided to reduce this mass. This hypothesis was proposed in~\cite{Torres:2000dw,Urena-Lopez:2002nup} and was further discussed in~\cite{Avilez:2017jql,Padilla:2020sjy,Lee:2015yws}, giving some optimistic results.

However, there is an interesting problem that we must face. Supermassive black holes at the center of galaxies could swallow all the SF. This problem was addressed in~\cite{Cruz-Osorio:2010nua} and years later in~\cite{Hui:2019aqm,Kiczek:2020gyd}. The results are that the SF can be accreted by the central SMBH, but at a very small rate, so that the SF can co-exist with the SMBH for longer than the lifetime of the universe.

  In 2023~\cite{2023arXiv231103412K} it was proposed that SFDM halos surrounding rotating SMBH binaries can emit dark matter waves through gravitational cooling. These waves could carry away orbital energy from the black holes, causing them to merge rapidly and potentially providing a solution to the final parsec problem. 

\section{Cosmological numerical simulations}

The first simulations of SFDM were performed at~\cite{Alcubierre:2001ea,Alcubierre:2002et,Alcubierre:2003sx}, where it was shown that SF collapse forms stable objects and can be compared to the halo of a galaxy, generating a central nucleus and with a natural cut in the mass power spectrum. However, the first 3D numerical simulations were performed at \cite{Schive:2014dra}, where previous results, such as the central nucleus in the center of the galaxy and the cutoff in the mass spectrum, were corroborated beyond doubt. This article started a series of works on 3D numerical simulations that were very successful; see, for example~\cite{Du:2016zcv,Du:2018qor,Veltmaat:2019hou,Church:2018sro,Davies:2019wgi,Mocz:2017wlg,Mocz:2019pyf,Mocz:2019uyd}, by reproducing the shape of the universe, we observe with fewer satellite galaxies and a central nucleus in galaxies without additional physics. Today, these numerical simulations can be compared with the same ones made with $\Lambda$CDM, showing only some essential differences. The most important thing is that, while $\Lambda$CDM predicts a central cusp density profile, SFDM predicts a galactic center, with a flat region called the soliton, discovered in 3D simulations in~\cite{Schive:2014dra}. This feature can make the difference between these two models. The other is the number of satellite galaxies around their hosts. While $\Lambda$CDM predicts thousands of satellite dark halos that need to be detected somehow, SFDM predicts a moderate number of them, which agree well with what is observed. Another feature that should be the difference is that the SFDM predicts that this soliton at the center of galaxies moves with time~\cite{Chowdhury:2021zik}, this feature is a footprint to be observed in galaxies if the SFDM is the DM in the universe.

The first numerical simulations with spherical symmetry also showed another problem; after the formation of the SFDM the object continues to collapse forming an object too dense to simulate a galaxy halo. This collapse is scale dependent; for large objects, such as galaxy clusters, this collapse takes so long that the formation of galaxy clusters can be explained very well with this model, but the formation of galaxies and dwarf galaxies collapse too soon~\cite{Guzman:2003kt,Guzman:2004wj,Guzman:2006yc}.
This problem was addressed for the first time in the literature considering the quantum characteristics of the scalar field using the excited states of the system in~\cite{Urena-Lopez:2010zva,Matos:2011pd}. The idea is that since the SFDM is a quantum-mechanical system, it should contain excited states that could give the system some additional stability. The main result here is that galaxy halos, even for small galaxies such as LSB or drwaf, remain stable~\cite{Urena-Lopez:2010zva,Guzman:2019gqc}. Therefore, it is necessary for the halo of galaxies to have at least two cohabiting states to adapt to the rotation curves of the galaxies.
With this in mind, in~\cite{Robles:2013dfa} some galaxy rotation curves were fitted using excited states of the SFDM, giving a good result. More recently, this idea has been used to explain other phenomena such as VPOs and Fermi bubbles, which we will discuss in the following.
This idea started a new paradigm in the literature called $\ell$-boson stars~\cite{Alcubierre:2018ahf}.

\section{The Vast Polar Orbits (VPO) problem}

For some time, astronomers have observed that the satellite galaxies of the Milky Way are not distributed homogeneously but that there are anomalous trajectories of the satellites of this galaxy, called VPO~\cite{Pawlowski:2013kpa,Pawlowski:2019bar}, establishing that among the 50 satellites in the Local Group, 43 are contained in four different planes, but this is inconsistent with simulations based on CDM because it predicts that this distribution must be isotropic. Furthermore, this alignment has also been observed in M31~\cite{Conn:2013iu,Ibata:2013rh}. Numerical CDM simulations predict that satellite galaxies around their host should be homogeneously distributed~\cite{Pawlowski:2018sys,Shaya:2013xna}. 
More recently, this same alignment of the paths of satellite galaxies has been observed in Centaurus, where 31 satellite galaxies in the constellation of Centaurus interact gravitationally with the elliptical galaxy Centaurus A and display a similar anisotropic alignment~\cite{Muller:2018hks}.
In a recent work~\cite{Solis-Lopez:2019lvz}, the authors took into account these excited states of the galaxy's halo, treating the galaxy's halo as an atom, to explain these observations and show that this behavior of the satellites is very natural in SFDM. This is so because the quantum character of the SFDM~\cite{Matos:2022rvo} allows us to put the states of the SF as in an atom. The ground state is spherically symmetric, but the first excited state has a figure-8 shape, aligning surrounding objects with the figure-8 shape. This is an important feature of the SFDM, the probability of finding this alignment is too low to be able to be explained as a simple chance, but if the SFDM is the real nature of the DM, we must find this alignment in more galaxies. These observations may be made in the near future.
In \cite{Park:2022lel} another mechanism for the formation of the satellite plane based on the gravitational cooling of SFDM was proposed.
\section{Detection}

It is actually very difficult to detect or design an experiment to detect a spinless particle with an ultralight mass. There are several hypotheses and proposals that attempt to do that, but so far these attempts have had limited results. We want to mention only some of them here.

In~\cite{Bozek:2014uqa} the authors use the MPS cutoff for the SFDM to construct a UV luminosity function and compare it to the Hubble ultra-deep field UV luminosity function, giving a constraint on the SFDM mass $m\geq10^{-22}$eV. 
Observables of 21 cm and fluctuations in CMB are proposed in~\cite{Kadota:2013iya}, including measurements of the CMB lens that can provide information on the existence of SFDM for different masses $m<10^{-26}$eV. 

Other possible ways to detect SFDM are using atomic methods such as hyperfine frequencies~\cite{Hees:2016gop}, 
or atomic clocks~\cite{Kouvaris:2019nzd,Filzinger:2023zrs}, 
interferometers~\cite{Aiello:2021wlp,Zhao:2021tie,Kim:2023pkx}, 
atom multigradiometry~\cite{Badurina:2022ngn,Badurina:2022ngn}, 
or neutrino interactions~\cite{Cordero:2022fwb}.

Gravitational waves produced by Black Hole (BH) mergers are one of the most anticipated ways to detect any sign of the existence of SFDM, e.g. gravitational waves emitted by BH~\cite{DAntonio:2018sff,Isi:2018pzk,Morisaki:2018htj,Palomba:2019vxe,Sun:2019mqb,Ng:2020jqd,Ng:2020ruv,Banerjee:2022zii,Liu:2021xfb,Manita:2023mnc,Yu:2023iog,Miller:2023kkd,Tsutsui:2023jbk},
or gravitational wave resonance~\cite{Delgado:2023psl},
using superradiance from BH~\cite{Hannuksela:2018izj},
BH mergers~\cite{Ng:2019jsx,Chung:2021roh,Chan:2022dkt}, or the deflection angle of black holes~\cite{Pantig:2022toh}. Another possible way to detect SFDM is observations in galaxies and their surroundings, e.g.,
with short-range gravity experiments~\cite{Qin:2022qnk},
or dynamical friction~\cite{Traykova:2021dua,Wang:2021udl,Vicente:2022ivh,Boudon:2022dxi}, 
binary pulsars~\cite{Blas:2019hxz}
compact eccentric binaries~\cite{Su:2021dwz}, 
extreme mass-ratio inspirals~\cite{Barsanti:2022vvl}, 
direct BH observations using the Event Horizon Telescope~\cite{Davoudiasl:2019nlo}, 
or by asteroid date~\cite{Tsai:2021irw,Chakrabarti:2022owq,Tsai:2023zza},
quadruply-imaged quasars~\cite{Laroche:2022pjm}
frame draggin effect~\cite{Poddar:2021ose}, 
kinetic Sunyaev-Zel’dovich eﬀect~\cite{Farren:2021jcd}, 
Shapiro delay~\cite{Poddar:2021sbc}, 
or in the center of our galaxy by
motion of the S2 star around Sgr A~\cite{DellaMonica:2023dcw,Yuan:2022nmu}.

An alternative is to use the Pulsar Timing Array (PTA) to detect SFDM particles~\cite{Porayko:2018sfa,NANOGrav:2023hvm,EPTA:2023xxk,Hwang:2023odi,EuropeanPulsarTimingArray:2023egv,Xia:2023hov}.

The existence of a central soliton is another distinguishing feature of SFDM~\cite{DeMartino:2018zkx}. Another interesting method is by tidal effects and tunneling out of satellite galaxies \cite{Dave:2023egr}. The number of proposals is huge, there are many possible ways to find a signal of the existence of SFDM, and it is possible that the detection of this type of DM is close.

\section{Conclusions and Challenges}
The SFDM model has proven to be a paradigm that is very capable of explaining DM in the universe. This model has surpassed 25 years of most new observations of the universe and every time the observations go further with better resolution, the model fits the observations with better clarity, especially cosmological observations which are now very high resolution. 
This model is today a true competitor to the cold dark matter model, as it offers alternative explanations for the phenomena we see in the universe, with fewer hypotheses and less additional physics. 

We believe that there are at least two observations that other DM models cannot explain, the first is that all satellite galaxies in our neighborhood contain the same amount of matter within the first 300pc. As we saw, the SFDM model is capable of explaining this fact in a very natural way. The other observation that can be explained using the SFDM in a very natural way is the VPOs.
One of the main characteristics of the SFDM is its quantum character \cite{Matos:2022rvo}, even though the SF is considered classical in this model, the SF complies with the Schr\"odinger equation that contains excited states. This is a characteristic unique to this model, therefore, if we see more galaxies with VPO or with Femi bubbles, this will be a strong corroboration that SFDM or some very similar candidate is the final answer for the DM nature in the universe. We do not know of any other model that can explain these observations in a way as natural as the SFDM model does.

Nevertheless, the model still faces some challenges, some of which we list here.
Perhaps one of the strongest constraints placed on the SFDM is that of the Lyman-$\alpha$ forest observations. In \cite{Irsic:2017yje,Armengaud:2017nkf,Nori:2018pka} it was found that the SFDM masses $m<2.3\times10^{-21}$eV are discarded. Of course, there is a possibility that the observations of the Lyman-$\alpha$ forest are not so fine that we cannot say with enough certainty that the model is incorrect; we have to expect new observations of this in the near future, e.g. DESI observations, which will come to light shortly. On the other hand, there are ways to solve this problem; one is provided in~\cite{Schive:2017biq} with an extension to the SFDM model or it has been argued in~\cite{Robles:2012kt} that the SFDM mass that we read in galaxies is the effective mass with the finite-temperature contribution, but the SFDM mass could be $m_\Phi\sim10^{-21}$eV, and each galaxy could show an apparently different mass because we read the effective mass due to the finite temperature of the SFDM in each galaxy. The temperature of the SFDM alters the scale $l$ for each galaxy as $l^2=\omega^2-\frac{\lambda}{2}\left(T^2_c-T^2\right)a^2l^2=\omega^2-m^2a^2$, explaining why we see that the mass of the SFDM at cosmological scales is $m_\Phi\sim10^{-21}$ eV and in galaxies could be $m\sim10^{-22}-10^{-24}$eV.
Incorporating the self-interaction of SFDM could be another potential solution to this problem.

The SFDM is today one of the most studied DM models in the community and it is not far from this model proving to be the last answer to the DM problem in the universe.

\section{Acknowledgments}
This work was partially supported by Programa para el Desarrollo Profesional Docente; Dirección de Apoyo a la Investigación y al Posgrado, Universidad de Guanajuato; CONACyT México under Grants No. A1-S-17899, A1-S-8742, 304001, 376127, 240512,  FORDECYT-PRONACES grant No. 490769 and I0101/131/07 C-234/07 of the Instituto Avanzado de Cosmolog\'ia (IAC) collaboration (http://www.iac.edu.mx/).


\begin{thebibliography}{140}
\expandafter\ifx\csname natexlab\endcsname\relax\def\natexlab#1{#1}\fi
\expandafter\ifx\csname bibnamefont\endcsname\relax
  \def\bibnamefont#1{#1}\fi
\expandafter\ifx\csname bibfnamefont\endcsname\relax
  \def\bibfnamefont#1{#1}\fi
\expandafter\ifx\csname citenamefont\endcsname\relax
  \def\citenamefont#1{#1}\fi
\expandafter\ifx\csname url\endcsname\relax
  \def\url#1{\texttt{#1}}\fi
\expandafter\ifx\csname urlprefix\endcsname\relax\def\urlprefix{URL }\fi
\providecommand{\bibinfo}[2]{#2}
\providecommand{\eprint}[2][]{\url{#2}}

\bibitem[{\citenamefont{Turner}(1983)}]{Turner:1983he}
\bibinfo{author}{\bibfnamefont{M.~S.} \bibnamefont{Turner}},
  \bibinfo{journal}{Phys. Rev. D} \textbf{\bibinfo{volume}{28}},
  \bibinfo{pages}{1243} (\bibinfo{year}{1983}).

\bibitem[{\citenamefont{Matos and Urena-Lopez}(2001)}]{Matos:2000ss}
\bibinfo{author}{\bibfnamefont{T.}~\bibnamefont{Matos}} \bibnamefont{and}
  \bibinfo{author}{\bibfnamefont{L.~A.} \bibnamefont{Urena-Lopez}},
  \bibinfo{journal}{Phys. Rev. D} \textbf{\bibinfo{volume}{63}},
  \bibinfo{pages}{063506} (\bibinfo{year}{2001}), \eprint{astro-ph/0006024}.

\bibitem[{\citenamefont{Sahni and Wang}(2000)}]{Sahni:1999qe}
\bibinfo{author}{\bibfnamefont{V.}~\bibnamefont{Sahni}} \bibnamefont{and}
  \bibinfo{author}{\bibfnamefont{L.-M.} \bibnamefont{Wang}},
  \bibinfo{journal}{Phys. Rev. D} \textbf{\bibinfo{volume}{62}},
  \bibinfo{pages}{103517} (\bibinfo{year}{2000}), \eprint{astro-ph/9910097}.

\bibitem[{\citenamefont{Baldeschi et~al.}(1983)\citenamefont{Baldeschi,
  Ruffini, and Gelmini}}]{Baldeschi:1983mq}
\bibinfo{author}{\bibfnamefont{M.~R.} \bibnamefont{Baldeschi}},
  \bibinfo{author}{\bibfnamefont{R.}~\bibnamefont{Ruffini}}, \bibnamefont{and}
  \bibinfo{author}{\bibfnamefont{G.~B.} \bibnamefont{Gelmini}},
  \bibinfo{journal}{Phys. Lett. B} \textbf{\bibinfo{volume}{122}},
  \bibinfo{pages}{221} (\bibinfo{year}{1983}).

\bibitem[{\citenamefont{Press et~al.}(1990)\citenamefont{Press, Ryden, and
  Spergel}}]{Press:1989id}
\bibinfo{author}{\bibfnamefont{W.~H.} \bibnamefont{Press}},
  \bibinfo{author}{\bibfnamefont{B.~S.} \bibnamefont{Ryden}}, \bibnamefont{and}
  \bibinfo{author}{\bibfnamefont{D.~N.} \bibnamefont{Spergel}},
  \bibinfo{journal}{Phys. Rev. Lett.} \textbf{\bibinfo{volume}{64}},
  \bibinfo{pages}{1084} (\bibinfo{year}{1990}).

\bibitem[{\citenamefont{Sin}(1994)}]{Sin:1992bg}
\bibinfo{author}{\bibfnamefont{S.-J.} \bibnamefont{Sin}},
  \bibinfo{journal}{Phys. Rev. D} \textbf{\bibinfo{volume}{50}},
  \bibinfo{pages}{3650} (\bibinfo{year}{1994}), \eprint{hep-ph/9205208}.

\bibitem[{\citenamefont{Ji and Sin}(1994)}]{Ji:1994xh}
\bibinfo{author}{\bibfnamefont{S.~U.} \bibnamefont{Ji}} \bibnamefont{and}
  \bibinfo{author}{\bibfnamefont{S.~J.} \bibnamefont{Sin}},
  \bibinfo{journal}{Phys. Rev. D} \textbf{\bibinfo{volume}{50}},
  \bibinfo{pages}{3655} (\bibinfo{year}{1994}), \eprint{hep-ph/9409267}.

\bibitem[{\citenamefont{Widrow and Kaiser}(1993)}]{Widrow:1993qq}
\bibinfo{author}{\bibfnamefont{L.~M.} \bibnamefont{Widrow}} \bibnamefont{and}
  \bibinfo{author}{\bibfnamefont{N.}~\bibnamefont{Kaiser}},
  \bibinfo{journal}{Astrophys. J. Lett.} \textbf{\bibinfo{volume}{416}},
  \bibinfo{pages}{L71} (\bibinfo{year}{1993}).

\bibitem[{\citenamefont{Lee and Koh}(1996)}]{Lee:1995af}
\bibinfo{author}{\bibfnamefont{J.-W.} \bibnamefont{Lee}} \bibnamefont{and}
  \bibinfo{author}{\bibfnamefont{I.-G.} \bibnamefont{Koh}},
  \bibinfo{journal}{Phys. Rev. D} \textbf{\bibinfo{volume}{53}},
  \bibinfo{pages}{2236} (\bibinfo{year}{1996}), \eprint{hep-ph/9507385}.

\bibitem[{\citenamefont{Matos and Guzman}(2000)}]{Matos:1998vk}
\bibinfo{author}{\bibfnamefont{T.}~\bibnamefont{Matos}} \bibnamefont{and}
  \bibinfo{author}{\bibfnamefont{F.~S.} \bibnamefont{Guzman}},
  \bibinfo{journal}{Class. Quant. Grav.} \textbf{\bibinfo{volume}{17}},
  \bibinfo{pages}{L9} (\bibinfo{year}{2000}), \eprint{gr-qc/9810028}.

\bibitem[{\citenamefont{Hu et~al.}(2000)\citenamefont{Hu, Barkana, and
  Gruzinov}}]{Hu:2000ke}
\bibinfo{author}{\bibfnamefont{W.}~\bibnamefont{Hu}},
  \bibinfo{author}{\bibfnamefont{R.}~\bibnamefont{Barkana}}, \bibnamefont{and}
  \bibinfo{author}{\bibfnamefont{A.}~\bibnamefont{Gruzinov}},
  \bibinfo{journal}{Phys. Rev. Lett.} \textbf{\bibinfo{volume}{85}},
  \bibinfo{pages}{1158} (\bibinfo{year}{2000}), \eprint{astro-ph/0003365}.

\bibitem[{\citenamefont{Arbey et~al.}(2001)\citenamefont{Arbey, Lesgourgues,
  and Salati}}]{Arbey:2001qi}
\bibinfo{author}{\bibfnamefont{A.}~\bibnamefont{Arbey}},
  \bibinfo{author}{\bibfnamefont{J.}~\bibnamefont{Lesgourgues}},
  \bibnamefont{and} \bibinfo{author}{\bibfnamefont{P.}~\bibnamefont{Salati}},
  \bibinfo{journal}{Phys. Rev. D} \textbf{\bibinfo{volume}{64}},
  \bibinfo{pages}{123528} (\bibinfo{year}{2001}), \eprint{astro-ph/0105564}.

\bibitem[{\citenamefont{Arbey et~al.}(2002)\citenamefont{Arbey, Lesgourgues,
  and Salati}}]{Arbey:2001jj}
\bibinfo{author}{\bibfnamefont{A.}~\bibnamefont{Arbey}},
  \bibinfo{author}{\bibfnamefont{J.}~\bibnamefont{Lesgourgues}},
  \bibnamefont{and} \bibinfo{author}{\bibfnamefont{P.}~\bibnamefont{Salati}},
  \bibinfo{journal}{Phys. Rev. D} \textbf{\bibinfo{volume}{65}},
  \bibinfo{pages}{083514} (\bibinfo{year}{2002}), \eprint{astro-ph/0112324}.

\bibitem[{\citenamefont{Amendola and Barbieri}(2006)}]{Amendola:2005ad}
\bibinfo{author}{\bibfnamefont{L.}~\bibnamefont{Amendola}} \bibnamefont{and}
  \bibinfo{author}{\bibfnamefont{R.}~\bibnamefont{Barbieri}},
  \bibinfo{journal}{Phys. Lett. B} \textbf{\bibinfo{volume}{642}},
  \bibinfo{pages}{192} (\bibinfo{year}{2006}), \eprint{hep-ph/0509257}.

\bibitem[{\citenamefont{Lundgren et~al.}(2010)\citenamefont{Lundgren,
  Bondarescu, Bondarescu, and Balakrishna}}]{Lundgren:2010sp}
\bibinfo{author}{\bibfnamefont{A.~P.} \bibnamefont{Lundgren}},
  \bibinfo{author}{\bibfnamefont{M.}~\bibnamefont{Bondarescu}},
  \bibinfo{author}{\bibfnamefont{R.}~\bibnamefont{Bondarescu}},
  \bibnamefont{and}
  \bibinfo{author}{\bibfnamefont{J.}~\bibnamefont{Balakrishna}},
  \bibinfo{journal}{Astrophys. J. Lett.} \textbf{\bibinfo{volume}{715}},
  \bibinfo{pages}{L35} (\bibinfo{year}{2010}), \eprint{1001.0051}.

\bibitem[{\citenamefont{Boehmer and Harko}(2007)}]{Boehmer:2007um}
\bibinfo{author}{\bibfnamefont{C.~G.} \bibnamefont{Boehmer}} \bibnamefont{and}
  \bibinfo{author}{\bibfnamefont{T.}~\bibnamefont{Harko}},
  \bibinfo{journal}{JCAP} \textbf{\bibinfo{volume}{06}}, \bibinfo{pages}{025}
  (\bibinfo{year}{2007}), \eprint{0705.4158}.

\bibitem[{\citenamefont{Rindler-Daller and
  Shapiro}(2010)}]{Rindler-Daller:2009qyu}
\bibinfo{author}{\bibfnamefont{T.}~\bibnamefont{Rindler-Daller}}
  \bibnamefont{and} \bibinfo{author}{\bibfnamefont{P.~R.}
  \bibnamefont{Shapiro}}, \bibinfo{journal}{ASP Conf. Ser.}
  \textbf{\bibinfo{volume}{432}}, \bibinfo{pages}{244} (\bibinfo{year}{2010}),
  \eprint{0912.2897}.

\bibitem[{\citenamefont{Chavanis}(2011)}]{Chavanis:2011gm}
\bibinfo{author}{\bibfnamefont{P.-H.} \bibnamefont{Chavanis}},
  \bibinfo{journal}{Phys. Rev. D} \textbf{\bibinfo{volume}{84}},
  \bibinfo{pages}{063518} (\bibinfo{year}{2011}), \eprint{1103.3219}.

\bibitem[{\citenamefont{Bray}(2010)}]{Bray:2010fc}
\bibinfo{author}{\bibfnamefont{H.~L.} \bibnamefont{Bray}}
  (\bibinfo{year}{2010}), \eprint{1004.4016}.

\bibitem[{\citenamefont{Schive et~al.}(2014)\citenamefont{Schive, Chiueh, and
  Broadhurst}}]{Schive:2014dra}
\bibinfo{author}{\bibfnamefont{H.-Y.} \bibnamefont{Schive}},
  \bibinfo{author}{\bibfnamefont{T.}~\bibnamefont{Chiueh}}, \bibnamefont{and}
  \bibinfo{author}{\bibfnamefont{T.}~\bibnamefont{Broadhurst}},
  \bibinfo{journal}{Nature Phys.} \textbf{\bibinfo{volume}{10}},
  \bibinfo{pages}{496} (\bibinfo{year}{2014}), \eprint{1406.6586}.

\bibitem[{\citenamefont{Hui et~al.}(2017)\citenamefont{Hui, Ostriker, Tremaine,
  and Witten}}]{Hui:2016ltb}
\bibinfo{author}{\bibfnamefont{L.}~\bibnamefont{Hui}},
  \bibinfo{author}{\bibfnamefont{J.~P.} \bibnamefont{Ostriker}},
  \bibinfo{author}{\bibfnamefont{S.}~\bibnamefont{Tremaine}}, \bibnamefont{and}
  \bibinfo{author}{\bibfnamefont{E.}~\bibnamefont{Witten}},
  \bibinfo{journal}{Phys. Rev. D} \textbf{\bibinfo{volume}{95}},
  \bibinfo{pages}{043541} (\bibinfo{year}{2017}), \eprint{1610.08297}.

\bibitem[{\citenamefont{Guzman and Urena-Lopez}(2003)}]{Guzman:2003kt}
\bibinfo{author}{\bibfnamefont{F.~S.} \bibnamefont{Guzman}} \bibnamefont{and}
  \bibinfo{author}{\bibfnamefont{L.~A.} \bibnamefont{Urena-Lopez}},
  \bibinfo{journal}{Phys. Rev. D} \textbf{\bibinfo{volume}{68}},
  \bibinfo{pages}{024023} (\bibinfo{year}{2003}), \eprint{astro-ph/0303440}.

\bibitem[{\citenamefont{Guzman and Urena-Lopez}(2004)}]{Guzman:2004wj}
\bibinfo{author}{\bibfnamefont{F.~S.} \bibnamefont{Guzman}} \bibnamefont{and}
  \bibinfo{author}{\bibfnamefont{L.~A.} \bibnamefont{Urena-Lopez}},
  \bibinfo{journal}{Phys. Rev. D} \textbf{\bibinfo{volume}{69}},
  \bibinfo{pages}{124033} (\bibinfo{year}{2004}), \eprint{gr-qc/0404014}.

\bibitem[{\citenamefont{Guzman and Urena-Lopez}(2006)}]{Guzman:2006yc}
\bibinfo{author}{\bibfnamefont{F.~S.} \bibnamefont{Guzman}} \bibnamefont{and}
  \bibinfo{author}{\bibfnamefont{L.~A.} \bibnamefont{Urena-Lopez}},
  \bibinfo{journal}{Astrophys. J.} \textbf{\bibinfo{volume}{645}},
  \bibinfo{pages}{814} (\bibinfo{year}{2006}), \eprint{astro-ph/0603613}.

\bibitem[{\citenamefont{Bray and Goetz}(2014)}]{Bray:2014dca}
\bibinfo{author}{\bibfnamefont{H.~L.} \bibnamefont{Bray}} \bibnamefont{and}
  \bibinfo{author}{\bibfnamefont{A.~S.} \bibnamefont{Goetz}}
  (\bibinfo{year}{2014}), \eprint{1409.7347}.

\bibitem[{\citenamefont{Lee et~al.}(2019)\citenamefont{Lee, Kim, and
  Lee}}]{Lee:2019ums}
\bibinfo{author}{\bibfnamefont{J.-W.} \bibnamefont{Lee}},
  \bibinfo{author}{\bibfnamefont{H.-C.} \bibnamefont{Kim}}, \bibnamefont{and}
  \bibinfo{author}{\bibfnamefont{J.}~\bibnamefont{Lee}},
  \bibinfo{journal}{Phys. Lett. B} \textbf{\bibinfo{volume}{795}},
  \bibinfo{pages}{206} (\bibinfo{year}{2019}), \eprint{1901.00305}.

\bibitem[{\citenamefont{Matos and Suarez}(2011)}]{Matos:2011kn}
\bibinfo{author}{\bibfnamefont{T.}~\bibnamefont{Matos}} \bibnamefont{and}
  \bibinfo{author}{\bibfnamefont{A.}~\bibnamefont{Suarez}},
  \bibinfo{journal}{EPL} \textbf{\bibinfo{volume}{96}}, \bibinfo{pages}{56005}
  (\bibinfo{year}{2011}), \eprint{1110.3114}.

\bibitem[{\citenamefont{Harko and Madarassy}(2012)}]{Harko:2011dz}
\bibinfo{author}{\bibfnamefont{T.}~\bibnamefont{Harko}} \bibnamefont{and}
  \bibinfo{author}{\bibfnamefont{E.~J.~M.} \bibnamefont{Madarassy}},
  \bibinfo{journal}{JCAP} \textbf{\bibinfo{volume}{01}}, \bibinfo{pages}{020}
  (\bibinfo{year}{2012}), \eprint{1110.2829}.

\bibitem[{\citenamefont{Matos and Su\'arez}(2014)}]{Matos:2011pd}
\bibinfo{author}{\bibfnamefont{T.}~\bibnamefont{Matos}} \bibnamefont{and}
  \bibinfo{author}{\bibfnamefont{A.}~\bibnamefont{Su\'arez}},
  \bibinfo{journal}{Class. Quant. Grav.} \textbf{\bibinfo{volume}{31}},
  \bibinfo{pages}{045015} (\bibinfo{year}{2014}), \eprint{1103.5731}.

\bibitem[{\citenamefont{Robles and Matos}(2013{\natexlab{a}})}]{Robles:2013dfa}
\bibinfo{author}{\bibfnamefont{V.~H.} \bibnamefont{Robles}} \bibnamefont{and}
  \bibinfo{author}{\bibfnamefont{T.}~\bibnamefont{Matos}},
  \bibinfo{journal}{Phys. Rev. D} \textbf{\bibinfo{volume}{88}},
  \bibinfo{pages}{083008} (\bibinfo{year}{2013}{\natexlab{a}}),
  \eprint{1302.5944}.

\bibitem[{\citenamefont{Harko}(2011{\natexlab{a}})}]{Harko:2011jy}
\bibinfo{author}{\bibfnamefont{T.}~\bibnamefont{Harko}}, \bibinfo{journal}{Mon.
  Not. Roy. Astron. Soc.} \textbf{\bibinfo{volume}{413}}, \bibinfo{pages}{3095}
  (\bibinfo{year}{2011}{\natexlab{a}}), \eprint{1101.3655}.

\bibitem[{\citenamefont{Ure\~na L\'opez and
  Gonzalez-Morales}(2016)}]{Urena-Lopez:2015gur}
\bibinfo{author}{\bibfnamefont{L.~A.} \bibnamefont{Ure\~na L\'opez}}
  \bibnamefont{and} \bibinfo{author}{\bibfnamefont{A.~X.}
  \bibnamefont{Gonzalez-Morales}}, \bibinfo{journal}{JCAP}
  \textbf{\bibinfo{volume}{07}}, \bibinfo{pages}{048} (\bibinfo{year}{2016}),
  \eprint{1511.08195}.

\bibitem[{\citenamefont{Hlozek et~al.}(2015)\citenamefont{Hlozek, Grin, Marsh,
  and Ferreira}}]{Hlozek:2014lca}
\bibinfo{author}{\bibfnamefont{R.}~\bibnamefont{Hlozek}},
  \bibinfo{author}{\bibfnamefont{D.}~\bibnamefont{Grin}},
  \bibinfo{author}{\bibfnamefont{D.~J.~E.} \bibnamefont{Marsh}},
  \bibnamefont{and} \bibinfo{author}{\bibfnamefont{P.~G.}
  \bibnamefont{Ferreira}}, \bibinfo{journal}{Phys. Rev. D}
  \textbf{\bibinfo{volume}{91}}, \bibinfo{pages}{103512}
  (\bibinfo{year}{2015}), \eprint{1410.2896}.

\bibitem[{\citenamefont{Cede\~no et~al.}(2017)\citenamefont{Cede\~no,
  Gonz\'alez-Morales, and Ure\~na L\'opez}}]{Cedeno:2017sou}
\bibinfo{author}{\bibfnamefont{F.~X.~L.} \bibnamefont{Cede\~no}},
  \bibinfo{author}{\bibfnamefont{A.~X.} \bibnamefont{Gonz\'alez-Morales}},
  \bibnamefont{and} \bibinfo{author}{\bibfnamefont{L.~A.} \bibnamefont{Ure\~na
  L\'opez}}, \bibinfo{journal}{Phys. Rev. D} \textbf{\bibinfo{volume}{96}},
  \bibinfo{pages}{061301} (\bibinfo{year}{2017}), \eprint{1703.10180}.

\bibitem[{\citenamefont{Li et~al.}(2014)\citenamefont{Li, Rindler-Daller, and
  Shapiro}}]{Li:2013nal}
\bibinfo{author}{\bibfnamefont{B.}~\bibnamefont{Li}},
  \bibinfo{author}{\bibfnamefont{T.}~\bibnamefont{Rindler-Daller}},
  \bibnamefont{and} \bibinfo{author}{\bibfnamefont{P.~R.}
  \bibnamefont{Shapiro}}, \bibinfo{journal}{Phys. Rev. D}
  \textbf{\bibinfo{volume}{89}}, \bibinfo{pages}{083536}
  (\bibinfo{year}{2014}), \eprint{1310.6061}.

\bibitem[{\citenamefont{Li et~al.}(2017)\citenamefont{Li, Shapiro, and
  Rindler-Daller}}]{Li:2016mmc}
\bibinfo{author}{\bibfnamefont{B.}~\bibnamefont{Li}},
  \bibinfo{author}{\bibfnamefont{P.~R.} \bibnamefont{Shapiro}},
  \bibnamefont{and}
  \bibinfo{author}{\bibfnamefont{T.}~\bibnamefont{Rindler-Daller}},
  \bibinfo{journal}{Phys. Rev. D} \textbf{\bibinfo{volume}{96}},
  \bibinfo{pages}{063505} (\bibinfo{year}{2017}), \eprint{1611.07961}.

\bibitem[{\citenamefont{Strigari et~al.}(2008)\citenamefont{Strigari, Bullock,
  Kaplinghat, Simon, Geha, Willman, and Walker}}]{Strigari:2008ib}
\bibinfo{author}{\bibfnamefont{L.~E.} \bibnamefont{Strigari}},
  \bibinfo{author}{\bibfnamefont{J.~S.} \bibnamefont{Bullock}},
  \bibinfo{author}{\bibfnamefont{M.}~\bibnamefont{Kaplinghat}},
  \bibinfo{author}{\bibfnamefont{J.~D.} \bibnamefont{Simon}},
  \bibinfo{author}{\bibfnamefont{M.}~\bibnamefont{Geha}},
  \bibinfo{author}{\bibfnamefont{B.}~\bibnamefont{Willman}}, \bibnamefont{and}
  \bibinfo{author}{\bibfnamefont{M.~G.} \bibnamefont{Walker}},
  \bibinfo{journal}{Nature} \textbf{\bibinfo{volume}{454}},
  \bibinfo{pages}{1096} (\bibinfo{year}{2008}), \eprint{0808.3772}.

\bibitem[{\citenamefont{Bernal et~al.}(2008)\citenamefont{Bernal, Matos, and
  N\'u\~nez}}]{Argelia08}
\bibinfo{author}{\bibfnamefont{A.}~\bibnamefont{Bernal}},
  \bibinfo{author}{\bibfnamefont{T.}~\bibnamefont{Matos}}, \bibnamefont{and}
  \bibinfo{author}{\bibfnamefont{D.}~\bibnamefont{N\'u\~nez}},
  \bibinfo{journal}{Revista Mexicana de Astronom\'ia y Astrof\'isica}
  \textbf{\bibinfo{volume}{44}}, \bibinfo{pages}{149} (\bibinfo{year}{2008}).

\bibitem[{\citenamefont{Su and Chen}(2011)}]{Su:2010bj}
\bibinfo{author}{\bibfnamefont{K.-Y.} \bibnamefont{Su}} \bibnamefont{and}
  \bibinfo{author}{\bibfnamefont{P.}~\bibnamefont{Chen}},
  \bibinfo{journal}{JCAP} \textbf{\bibinfo{volume}{08}}, \bibinfo{pages}{016}
  (\bibinfo{year}{2011}), \eprint{1008.3717}.

\bibitem[{\citenamefont{Harko}(2011{\natexlab{b}})}]{Harko:2011xw}
\bibinfo{author}{\bibfnamefont{T.}~\bibnamefont{Harko}},
  \bibinfo{journal}{JCAP} \textbf{\bibinfo{volume}{05}}, \bibinfo{pages}{022}
  (\bibinfo{year}{2011}{\natexlab{b}}), \eprint{1105.2996}.

\bibitem[{\citenamefont{Du et~al.}(2017)\citenamefont{Du, Behrens, and
  Niemeyer}}]{Du:2016zcv}
\bibinfo{author}{\bibfnamefont{X.}~\bibnamefont{Du}},
  \bibinfo{author}{\bibfnamefont{C.}~\bibnamefont{Behrens}}, \bibnamefont{and}
  \bibinfo{author}{\bibfnamefont{J.~C.} \bibnamefont{Niemeyer}},
  \bibinfo{journal}{Mon. Not. Roy. Astron. Soc.}
  \textbf{\bibinfo{volume}{465}}, \bibinfo{pages}{941} (\bibinfo{year}{2017}),
  \eprint{1608.02575}.

\bibitem[{\citenamefont{Mocz et~al.}(2019)}]{Mocz:2019pyf}
\bibinfo{author}{\bibfnamefont{P.}~\bibnamefont{Mocz}} \bibnamefont{et~al.},
  \bibinfo{journal}{Phys. Rev. Lett.} \textbf{\bibinfo{volume}{123}},
  \bibinfo{pages}{141301} (\bibinfo{year}{2019}), \eprint{1910.01653}.

\bibitem[{\citenamefont{Lee and Lim}(2010)}]{Lee:2008jp}
\bibinfo{author}{\bibfnamefont{J.-W.} \bibnamefont{Lee}} \bibnamefont{and}
  \bibinfo{author}{\bibfnamefont{S.}~\bibnamefont{Lim}},
  \bibinfo{journal}{JCAP} \textbf{\bibinfo{volume}{01}}, \bibinfo{pages}{007}
  (\bibinfo{year}{2010}), \eprint{0812.1342}.

\bibitem[{\citenamefont{Ure\~na L\'opez et~al.}(2017)\citenamefont{Ure\~na
  L\'opez, Robles, and Matos}}]{Urena-Lopez:2017tob}
\bibinfo{author}{\bibfnamefont{L.~A.} \bibnamefont{Ure\~na L\'opez}},
  \bibinfo{author}{\bibfnamefont{V.~H.} \bibnamefont{Robles}},
  \bibnamefont{and} \bibinfo{author}{\bibfnamefont{T.}~\bibnamefont{Matos}},
  \bibinfo{journal}{Phys. Rev. D} \textbf{\bibinfo{volume}{96}},
  \bibinfo{pages}{043005} (\bibinfo{year}{2017}), \eprint{1702.05103}.

\bibitem[{\citenamefont{Bozek et~al.}(2015)\citenamefont{Bozek, Marsh, Silk,
  and Wyse}}]{Bozek:2014uqa}
\bibinfo{author}{\bibfnamefont{B.}~\bibnamefont{Bozek}},
  \bibinfo{author}{\bibfnamefont{D.~J.~E.} \bibnamefont{Marsh}},
  \bibinfo{author}{\bibfnamefont{J.}~\bibnamefont{Silk}}, \bibnamefont{and}
  \bibinfo{author}{\bibfnamefont{R.~F.~G.} \bibnamefont{Wyse}},
  \bibinfo{journal}{Mon. Not. Roy. Astron. Soc.}
  \textbf{\bibinfo{volume}{450}}, \bibinfo{pages}{209} (\bibinfo{year}{2015}),
  \eprint{1409.3544}.

\bibitem[{\citenamefont{Lee}(2009)}]{Lee:2008ux}
\bibinfo{author}{\bibfnamefont{J.-W.} \bibnamefont{Lee}},
  \bibinfo{journal}{Phys. Lett. B} \textbf{\bibinfo{volume}{681}},
  \bibinfo{pages}{118} (\bibinfo{year}{2009}), \eprint{0805.2877}.

\bibitem[{\citenamefont{Lee}(2016)}]{Lee:2015cos}
\bibinfo{author}{\bibfnamefont{J.-W.} \bibnamefont{Lee}},
  \bibinfo{journal}{Phys. Lett. B} \textbf{\bibinfo{volume}{756}},
  \bibinfo{pages}{166} (\bibinfo{year}{2016}), \eprint{1511.06611}.

\bibitem[{\citenamefont{Seidel and Suen}(1991)}]{Seidel:1991zh}
\bibinfo{author}{\bibfnamefont{E.}~\bibnamefont{Seidel}} \bibnamefont{and}
  \bibinfo{author}{\bibfnamefont{W.~M.} \bibnamefont{Suen}},
  \bibinfo{journal}{Phys. Rev. Lett.} \textbf{\bibinfo{volume}{66}},
  \bibinfo{pages}{1659} (\bibinfo{year}{1991}).

\bibitem[{\citenamefont{Balakrishna et~al.}(1998)\citenamefont{Balakrishna,
  Seidel, and Suen}}]{Balakrishna:1997ej}
\bibinfo{author}{\bibfnamefont{J.}~\bibnamefont{Balakrishna}},
  \bibinfo{author}{\bibfnamefont{E.}~\bibnamefont{Seidel}}, \bibnamefont{and}
  \bibinfo{author}{\bibfnamefont{W.-M.} \bibnamefont{Suen}},
  \bibinfo{journal}{Phys. Rev. D} \textbf{\bibinfo{volume}{58}},
  \bibinfo{pages}{104004} (\bibinfo{year}{1998}), \eprint{gr-qc/9712064}.

\bibitem[{\citenamefont{Torres et~al.}(2000)\citenamefont{Torres, Capozziello,
  and Lambiase}}]{Torres:2000dw}
\bibinfo{author}{\bibfnamefont{D.~F.} \bibnamefont{Torres}},
  \bibinfo{author}{\bibfnamefont{S.}~\bibnamefont{Capozziello}},
  \bibnamefont{and} \bibinfo{author}{\bibfnamefont{G.}~\bibnamefont{Lambiase}},
  \bibinfo{journal}{Phys. Rev. D} \textbf{\bibinfo{volume}{62}},
  \bibinfo{pages}{104012} (\bibinfo{year}{2000}), \eprint{astro-ph/0004064}.

\bibitem[{\citenamefont{Urena-Lopez and Liddle}(2002)}]{Urena-Lopez:2002nup}
\bibinfo{author}{\bibfnamefont{L.~A.} \bibnamefont{Urena-Lopez}}
  \bibnamefont{and} \bibinfo{author}{\bibfnamefont{A.~R.}
  \bibnamefont{Liddle}}, \bibinfo{journal}{Phys. Rev. D}
  \textbf{\bibinfo{volume}{66}}, \bibinfo{pages}{083005}
  (\bibinfo{year}{2002}), \eprint{astro-ph/0207493}.

\bibitem[{\citenamefont{Avilez et~al.}(2018)\citenamefont{Avilez, Padilla,
  Bernal-Marin, and Matos}}]{Avilez:2017jql}
\bibinfo{author}{\bibfnamefont{A.~A.} \bibnamefont{Avilez}},
  \bibinfo{author}{\bibfnamefont{L.~E.} \bibnamefont{Padilla}},
  \bibinfo{author}{\bibfnamefont{T.}~\bibnamefont{Bernal-Marin}},
  \bibnamefont{and} \bibinfo{author}{\bibfnamefont{T.}~\bibnamefont{Matos}},
  \bibinfo{journal}{Mon. Not. Roy. Astron. Soc.}
  \textbf{\bibinfo{volume}{477}}, \bibinfo{pages}{3257} (\bibinfo{year}{2018}),
  \eprint{1704.07314}.

\bibitem[{\citenamefont{Padilla et~al.}(2021)\citenamefont{Padilla,
  Rindler-Daller, Shapiro, Matos, and V\'azquez}}]{Padilla:2020sjy}
\bibinfo{author}{\bibfnamefont{L.~E.} \bibnamefont{Padilla}},
  \bibinfo{author}{\bibfnamefont{T.}~\bibnamefont{Rindler-Daller}},
  \bibinfo{author}{\bibfnamefont{P.~R.} \bibnamefont{Shapiro}},
  \bibinfo{author}{\bibfnamefont{T.}~\bibnamefont{Matos}}, \bibnamefont{and}
  \bibinfo{author}{\bibfnamefont{J.~A.} \bibnamefont{V\'azquez}},
  \bibinfo{journal}{Phys. Rev. D} \textbf{\bibinfo{volume}{103}},
  \bibinfo{pages}{063012} (\bibinfo{year}{2021}), \eprint{2010.12716}.

\bibitem[{\citenamefont{Lee et~al.}(2020)\citenamefont{Lee, Lee, and
  Kim}}]{Lee:2015yws}
\bibinfo{author}{\bibfnamefont{J.-W.} \bibnamefont{Lee}},
  \bibinfo{author}{\bibfnamefont{J.}~\bibnamefont{Lee}}, \bibnamefont{and}
  \bibinfo{author}{\bibfnamefont{H.-C.} \bibnamefont{Kim}},
  \bibinfo{journal}{Mod. Phys. Lett. A} \textbf{\bibinfo{volume}{35}},
  \bibinfo{pages}{2050155} (\bibinfo{year}{2020}), \eprint{1512.02351}.

\bibitem[{\citenamefont{Cruz-Osorio et~al.}(2011)\citenamefont{Cruz-Osorio,
  Guzman, and Lora-Clavijo}}]{Cruz-Osorio:2010nua}
\bibinfo{author}{\bibfnamefont{A.}~\bibnamefont{Cruz-Osorio}},
  \bibinfo{author}{\bibfnamefont{F.~S.} \bibnamefont{Guzman}},
  \bibnamefont{and} \bibinfo{author}{\bibfnamefont{F.~D.}
  \bibnamefont{Lora-Clavijo}}, \bibinfo{journal}{JCAP}
  \textbf{\bibinfo{volume}{06}}, \bibinfo{pages}{029} (\bibinfo{year}{2011}),
  \eprint{1008.0027}.

\bibitem[{\citenamefont{Hui et~al.}(2019)\citenamefont{Hui, Kabat, Li, Santoni,
  and Wong}}]{Hui:2019aqm}
\bibinfo{author}{\bibfnamefont{L.}~\bibnamefont{Hui}},
  \bibinfo{author}{\bibfnamefont{D.}~\bibnamefont{Kabat}},
  \bibinfo{author}{\bibfnamefont{X.}~\bibnamefont{Li}},
  \bibinfo{author}{\bibfnamefont{L.}~\bibnamefont{Santoni}}, \bibnamefont{and}
  \bibinfo{author}{\bibfnamefont{S.~S.~C.} \bibnamefont{Wong}},
  \bibinfo{journal}{JCAP} \textbf{\bibinfo{volume}{06}}, \bibinfo{pages}{038}
  (\bibinfo{year}{2019}), \eprint{1904.12803}.

\bibitem[{\citenamefont{Kiczek and Rogatko}(2020)}]{Kiczek:2020gyd}
\bibinfo{author}{\bibfnamefont{B.}~\bibnamefont{Kiczek}} \bibnamefont{and}
  \bibinfo{author}{\bibfnamefont{M.}~\bibnamefont{Rogatko}},
  \bibinfo{journal}{Phys. Rev. D} \textbf{\bibinfo{volume}{101}},
  \bibinfo{pages}{084035} (\bibinfo{year}{2020}), \eprint{2004.06617}.

\bibitem[{\citenamefont{{Koo} et~al.}(2023)\citenamefont{{Koo}, {Bak}, {Park},
  {Hong}, and {Lee}}}]{2023arXiv231103412K}
\bibinfo{author}{\bibfnamefont{H.}~\bibnamefont{{Koo}}},
  \bibinfo{author}{\bibfnamefont{D.}~\bibnamefont{{Bak}}},
  \bibinfo{author}{\bibfnamefont{I.}~\bibnamefont{{Park}}},
  \bibinfo{author}{\bibfnamefont{S.~E.} \bibnamefont{{Hong}}},
  \bibnamefont{and} \bibinfo{author}{\bibfnamefont{J.-W.} \bibnamefont{{Lee}}},
  \bibinfo{journal}{arXiv e-prints} \bibinfo{eid}{arXiv:2311.03412}
  (\bibinfo{year}{2023}), \eprint{2311.03412}.

\bibitem[{\citenamefont{Alcubierre
  et~al.}(2002{\natexlab{a}})\citenamefont{Alcubierre, Guzman, Matos, Nunez,
  Urena-Lopez, and Wiederhold}}]{Alcubierre:2001ea}
\bibinfo{author}{\bibfnamefont{M.}~\bibnamefont{Alcubierre}},
  \bibinfo{author}{\bibfnamefont{F.~S.} \bibnamefont{Guzman}},
  \bibinfo{author}{\bibfnamefont{T.}~\bibnamefont{Matos}},
  \bibinfo{author}{\bibfnamefont{D.}~\bibnamefont{Nunez}},
  \bibinfo{author}{\bibfnamefont{L.~A.} \bibnamefont{Urena-Lopez}},
  \bibnamefont{and}
  \bibinfo{author}{\bibfnamefont{P.}~\bibnamefont{Wiederhold}},
  \bibinfo{journal}{Class. Quant. Grav.} \textbf{\bibinfo{volume}{19}},
  \bibinfo{pages}{5017} (\bibinfo{year}{2002}{\natexlab{a}}),
  \eprint{gr-qc/0110102}.

\bibitem[{\citenamefont{Alcubierre
  et~al.}(2002{\natexlab{b}})\citenamefont{Alcubierre, Guzman, Matos, Nunez,
  Urena-Lopez, and Wiederhold}}]{Alcubierre:2002et}
\bibinfo{author}{\bibfnamefont{M.}~\bibnamefont{Alcubierre}},
  \bibinfo{author}{\bibfnamefont{F.~S.} \bibnamefont{Guzman}},
  \bibinfo{author}{\bibfnamefont{T.}~\bibnamefont{Matos}},
  \bibinfo{author}{\bibfnamefont{D.}~\bibnamefont{Nunez}},
  \bibinfo{author}{\bibfnamefont{L.~A.} \bibnamefont{Urena-Lopez}},
  \bibnamefont{and}
  \bibinfo{author}{\bibfnamefont{P.}~\bibnamefont{Wiederhold}}, in
  \emph{\bibinfo{booktitle}{{4th International Heidelberg Conference on DM}}}
  (\bibinfo{year}{2002}{\natexlab{b}}), pp. \bibinfo{pages}{356--364},
  \eprint{astro-ph/0204307}.

\bibitem[{\citenamefont{Alcubierre et~al.}(2003)\citenamefont{Alcubierre,
  Becerril, Guzman, Matos, Nunez, and Urena-Lopez}}]{Alcubierre:2003sx}
\bibinfo{author}{\bibfnamefont{M.}~\bibnamefont{Alcubierre}},
  \bibinfo{author}{\bibfnamefont{R.}~\bibnamefont{Becerril}},
  \bibinfo{author}{\bibfnamefont{S.~F.} \bibnamefont{Guzman}},
  \bibinfo{author}{\bibfnamefont{T.}~\bibnamefont{Matos}},
  \bibinfo{author}{\bibfnamefont{D.}~\bibnamefont{Nunez}}, \bibnamefont{and}
  \bibinfo{author}{\bibfnamefont{L.~A.} \bibnamefont{Urena-Lopez}},
  \bibinfo{journal}{Class. Quant. Grav.} \textbf{\bibinfo{volume}{20}},
  \bibinfo{pages}{2883} (\bibinfo{year}{2003}), \eprint{gr-qc/0301105}.

\bibitem[{\citenamefont{Du et~al.}(2018)\citenamefont{Du, Schwabe, Niemeyer,
  and B\"urger}}]{Du:2018qor}
\bibinfo{author}{\bibfnamefont{X.}~\bibnamefont{Du}},
  \bibinfo{author}{\bibfnamefont{B.}~\bibnamefont{Schwabe}},
  \bibinfo{author}{\bibfnamefont{J.~C.} \bibnamefont{Niemeyer}},
  \bibnamefont{and} \bibinfo{author}{\bibfnamefont{D.}~\bibnamefont{B\"urger}},
  \bibinfo{journal}{Phys. Rev. D} \textbf{\bibinfo{volume}{97}},
  \bibinfo{pages}{063507} (\bibinfo{year}{2018}), \eprint{1801.04864}.

\bibitem[{\citenamefont{Veltmaat et~al.}(2020)\citenamefont{Veltmaat, Schwabe,
  and Niemeyer}}]{Veltmaat:2019hou}
\bibinfo{author}{\bibfnamefont{J.}~\bibnamefont{Veltmaat}},
  \bibinfo{author}{\bibfnamefont{B.}~\bibnamefont{Schwabe}}, \bibnamefont{and}
  \bibinfo{author}{\bibfnamefont{J.~C.} \bibnamefont{Niemeyer}},
  \bibinfo{journal}{Phys. Rev. D} \textbf{\bibinfo{volume}{101}},
  \bibinfo{pages}{083518} (\bibinfo{year}{2020}), \eprint{1911.09614}.

\bibitem[{\citenamefont{Church et~al.}(2019)\citenamefont{Church, Ostriker, and
  Mocz}}]{Church:2018sro}
\bibinfo{author}{\bibfnamefont{B.~V.} \bibnamefont{Church}},
  \bibinfo{author}{\bibfnamefont{J.~P.} \bibnamefont{Ostriker}},
  \bibnamefont{and} \bibinfo{author}{\bibfnamefont{P.}~\bibnamefont{Mocz}},
  \bibinfo{journal}{Mon. Not. Roy. Astron. Soc.}
  \textbf{\bibinfo{volume}{485}}, \bibinfo{pages}{2861} (\bibinfo{year}{2019}),
  \eprint{1809.04744}.

\bibitem[{\citenamefont{Davies and Mocz}(2020)}]{Davies:2019wgi}
\bibinfo{author}{\bibfnamefont{E.~Y.} \bibnamefont{Davies}} \bibnamefont{and}
  \bibinfo{author}{\bibfnamefont{P.}~\bibnamefont{Mocz}},
  \bibinfo{journal}{Mon. Not. Roy. Astron. Soc.}
  \textbf{\bibinfo{volume}{492}}, \bibinfo{pages}{5721} (\bibinfo{year}{2020}),
  \eprint{1908.04790}.

\bibitem[{\citenamefont{Mocz et~al.}(2017)\citenamefont{Mocz, Vogelsberger,
  Robles, Zavala, Boylan-Kolchin, Fialkov, and Hernquist}}]{Mocz:2017wlg}
\bibinfo{author}{\bibfnamefont{P.}~\bibnamefont{Mocz}},
  \bibinfo{author}{\bibfnamefont{M.}~\bibnamefont{Vogelsberger}},
  \bibinfo{author}{\bibfnamefont{V.~H.} \bibnamefont{Robles}},
  \bibinfo{author}{\bibfnamefont{J.}~\bibnamefont{Zavala}},
  \bibinfo{author}{\bibfnamefont{M.}~\bibnamefont{Boylan-Kolchin}},
  \bibinfo{author}{\bibfnamefont{A.}~\bibnamefont{Fialkov}}, \bibnamefont{and}
  \bibinfo{author}{\bibfnamefont{L.}~\bibnamefont{Hernquist}},
  \bibinfo{journal}{Mon. Not. Roy. Astron. Soc.}
  \textbf{\bibinfo{volume}{471}}, \bibinfo{pages}{4559} (\bibinfo{year}{2017}),
  \eprint{1705.05845}.

\bibitem[{\citenamefont{Mocz et~al.}(2020)}]{Mocz:2019uyd}
\bibinfo{author}{\bibfnamefont{P.}~\bibnamefont{Mocz}} \bibnamefont{et~al.},
  \bibinfo{journal}{Mon. Not. Roy. Astron. Soc.}
  \textbf{\bibinfo{volume}{494}}, \bibinfo{pages}{2027} (\bibinfo{year}{2020}),
  \eprint{1911.05746}.

\bibitem[{\citenamefont{Chowdhury et~al.}(2021)\citenamefont{Chowdhury, van~den
  Bosch, Robles, van Dokkum, Schive, Chiueh, and
  Broadhurst}}]{Chowdhury:2021zik}
\bibinfo{author}{\bibfnamefont{D.~D.} \bibnamefont{Chowdhury}},
  \bibinfo{author}{\bibfnamefont{F.~C.} \bibnamefont{van~den Bosch}},
  \bibinfo{author}{\bibfnamefont{V.~H.} \bibnamefont{Robles}},
  \bibinfo{author}{\bibfnamefont{P.}~\bibnamefont{van Dokkum}},
  \bibinfo{author}{\bibfnamefont{H.-Y.} \bibnamefont{Schive}},
  \bibinfo{author}{\bibfnamefont{T.}~\bibnamefont{Chiueh}}, \bibnamefont{and}
  \bibinfo{author}{\bibfnamefont{T.}~\bibnamefont{Broadhurst}},
  \bibinfo{journal}{Astrophys. J.} \textbf{\bibinfo{volume}{916}},
  \bibinfo{pages}{27} (\bibinfo{year}{2021}), \eprint{2105.05268}.

\bibitem[{\citenamefont{Urena-Lopez and Bernal}(2010)}]{Urena-Lopez:2010zva}
\bibinfo{author}{\bibfnamefont{L.~A.} \bibnamefont{Urena-Lopez}}
  \bibnamefont{and} \bibinfo{author}{\bibfnamefont{A.}~\bibnamefont{Bernal}},
  \bibinfo{journal}{Phys. Rev. D} \textbf{\bibinfo{volume}{82}},
  \bibinfo{pages}{123535} (\bibinfo{year}{2010}), \eprint{1008.1231}.

\bibitem[{\citenamefont{Guzm\'an and Ure\~na L\'opez}(2020)}]{Guzman:2019gqc}
\bibinfo{author}{\bibfnamefont{F.~S.} \bibnamefont{Guzm\'an}} \bibnamefont{and}
  \bibinfo{author}{\bibfnamefont{L.~A.} \bibnamefont{Ure\~na L\'opez}},
  \bibinfo{journal}{Phys. Rev. D} \textbf{\bibinfo{volume}{101}},
  \bibinfo{pages}{081302} (\bibinfo{year}{2020}), \eprint{1912.10585}.

\bibitem[{\citenamefont{Alcubierre et~al.}(2018)\citenamefont{Alcubierre,
  Barranco, Bernal, Degollado, Diez-Tejedor, Megevand, Nunez, and
  Sarbach}}]{Alcubierre:2018ahf}
\bibinfo{author}{\bibfnamefont{M.}~\bibnamefont{Alcubierre}},
  \bibinfo{author}{\bibfnamefont{J.}~\bibnamefont{Barranco}},
  \bibinfo{author}{\bibfnamefont{A.}~\bibnamefont{Bernal}},
  \bibinfo{author}{\bibfnamefont{J.~C.} \bibnamefont{Degollado}},
  \bibinfo{author}{\bibfnamefont{A.}~\bibnamefont{Diez-Tejedor}},
  \bibinfo{author}{\bibfnamefont{M.}~\bibnamefont{Megevand}},
  \bibinfo{author}{\bibfnamefont{D.}~\bibnamefont{Nunez}}, \bibnamefont{and}
  \bibinfo{author}{\bibfnamefont{O.}~\bibnamefont{Sarbach}},
  \bibinfo{journal}{Class. Quant. Grav.} \textbf{\bibinfo{volume}{35}},
  \bibinfo{pages}{19LT01} (\bibinfo{year}{2018}), \eprint{1805.11488}.

\bibitem[{\citenamefont{Pawlowski et~al.}(2013)\citenamefont{Pawlowski, Kroupa,
  and Jerjen}}]{Pawlowski:2013kpa}
\bibinfo{author}{\bibfnamefont{M.~S.} \bibnamefont{Pawlowski}},
  \bibinfo{author}{\bibfnamefont{P.}~\bibnamefont{Kroupa}}, \bibnamefont{and}
  \bibinfo{author}{\bibfnamefont{H.}~\bibnamefont{Jerjen}},
  \bibinfo{journal}{Mon. Not. Roy. Astron. Soc.}
  \textbf{\bibinfo{volume}{435}}, \bibinfo{pages}{1928} (\bibinfo{year}{2013}),
  \eprint{1307.6210}.

\bibitem[{\citenamefont{Pawlowski and Kroupa}(2020)}]{Pawlowski:2019bar}
\bibinfo{author}{\bibfnamefont{M.~S.} \bibnamefont{Pawlowski}}
  \bibnamefont{and} \bibinfo{author}{\bibfnamefont{P.}~\bibnamefont{Kroupa}},
  \bibinfo{journal}{Mon. Not. Roy. Astron. Soc.}
  \textbf{\bibinfo{volume}{491}}, \bibinfo{pages}{3042} (\bibinfo{year}{2020}),
  \eprint{1911.05081}.

\bibitem[{\citenamefont{Conn et~al.}(2013)}]{Conn:2013iu}
\bibinfo{author}{\bibfnamefont{A.~R.} \bibnamefont{Conn}} \bibnamefont{et~al.},
  \bibinfo{journal}{Astrophys. J.} \textbf{\bibinfo{volume}{766}},
  \bibinfo{pages}{120} (\bibinfo{year}{2013}), \eprint{1301.7131}.

\bibitem[{\citenamefont{Ibata et~al.}(2013)}]{Ibata:2013rh}
\bibinfo{author}{\bibfnamefont{R.~A.} \bibnamefont{Ibata}}
  \bibnamefont{et~al.}, \bibinfo{journal}{Nature}
  \textbf{\bibinfo{volume}{493}}, \bibinfo{pages}{62} (\bibinfo{year}{2013}),
  \eprint{1301.0446}.

\bibitem[{\citenamefont{Pawlowski}(2018)}]{Pawlowski:2018sys}
\bibinfo{author}{\bibfnamefont{M.~S.} \bibnamefont{Pawlowski}},
  \bibinfo{journal}{Mod. Phys. Lett. A} \textbf{\bibinfo{volume}{33}},
  \bibinfo{pages}{1830004} (\bibinfo{year}{2018}), \eprint{1802.02579}.

\bibitem[{\citenamefont{Shaya and Tully}(2013)}]{Shaya:2013xna}
\bibinfo{author}{\bibfnamefont{E.~J.} \bibnamefont{Shaya}} \bibnamefont{and}
  \bibinfo{author}{\bibfnamefont{R.~B.} \bibnamefont{Tully}},
  \bibinfo{journal}{Mon. Not. Roy. Astron. Soc.}
  \textbf{\bibinfo{volume}{436}}, \bibinfo{pages}{2096} (\bibinfo{year}{2013}),
  \eprint{1307.4297}.

\bibitem[{\citenamefont{M\"uller et~al.}(2018)\citenamefont{M\"uller,
  Pawlowski, Jerjen, and Lelli}}]{Muller:2018hks}
\bibinfo{author}{\bibfnamefont{O.}~\bibnamefont{M\"uller}},
  \bibinfo{author}{\bibfnamefont{M.~S.} \bibnamefont{Pawlowski}},
  \bibinfo{author}{\bibfnamefont{H.}~\bibnamefont{Jerjen}}, \bibnamefont{and}
  \bibinfo{author}{\bibfnamefont{F.}~\bibnamefont{Lelli}},
  \bibinfo{journal}{Science} \textbf{\bibinfo{volume}{359}},
  \bibinfo{pages}{534} (\bibinfo{year}{2018}), \eprint{1802.00081}.

\bibitem[{\citenamefont{Sol\'\i{}s-L\'opez
  et~al.}(2021)\citenamefont{Sol\'\i{}s-L\'opez, Guzm\'an, Matos, Robles, and
  Ure\~na L\'opez}}]{Solis-Lopez:2019lvz}
\bibinfo{author}{\bibfnamefont{J.}~\bibnamefont{Sol\'\i{}s-L\'opez}},
  \bibinfo{author}{\bibfnamefont{F.~S.} \bibnamefont{Guzm\'an}},
  \bibinfo{author}{\bibfnamefont{T.}~\bibnamefont{Matos}},
  \bibinfo{author}{\bibfnamefont{V.~H.} \bibnamefont{Robles}},
  \bibnamefont{and} \bibinfo{author}{\bibfnamefont{L.~A.} \bibnamefont{Ure\~na
  L\'opez}}, \bibinfo{journal}{Phys. Rev. D} \textbf{\bibinfo{volume}{103}},
  \bibinfo{pages}{083535} (\bibinfo{year}{2021}), \eprint{1912.09660}.

\bibitem[{\citenamefont{Matos}(2022)}]{Matos:2022rvo}
\bibinfo{author}{\bibfnamefont{T.}~\bibnamefont{Matos}}, \bibinfo{journal}{Mon.
  Not. Roy. Astron. Soc.} \textbf{\bibinfo{volume}{517}}, \bibinfo{pages}{5247}
  (\bibinfo{year}{2022}), \eprint{2211.02025}.

\bibitem[{\citenamefont{Park et~al.}(2022)\citenamefont{Park, Bak, Lee, and
  Park}}]{Park:2022lel}
\bibinfo{author}{\bibfnamefont{S.}~\bibnamefont{Park}},
  \bibinfo{author}{\bibfnamefont{D.}~\bibnamefont{Bak}},
  \bibinfo{author}{\bibfnamefont{J.-W.} \bibnamefont{Lee}}, \bibnamefont{and}
  \bibinfo{author}{\bibfnamefont{I.}~\bibnamefont{Park}},
  \bibinfo{journal}{JCAP} \textbf{\bibinfo{volume}{12}}, \bibinfo{pages}{033}
  (\bibinfo{year}{2022}), \eprint{2207.07192}.

\bibitem[{\citenamefont{Kadota et~al.}(2014)\citenamefont{Kadota, Mao, Ichiki,
  and Silk}}]{Kadota:2013iya}
\bibinfo{author}{\bibfnamefont{K.}~\bibnamefont{Kadota}},
  \bibinfo{author}{\bibfnamefont{Y.}~\bibnamefont{Mao}},
  \bibinfo{author}{\bibfnamefont{K.}~\bibnamefont{Ichiki}}, \bibnamefont{and}
  \bibinfo{author}{\bibfnamefont{J.}~\bibnamefont{Silk}},
  \bibinfo{journal}{JCAP} \textbf{\bibinfo{volume}{06}}, \bibinfo{pages}{011}
  (\bibinfo{year}{2014}), \eprint{1312.1898}.

\bibitem[{\citenamefont{Hees et~al.}(2016)\citenamefont{Hees, Gu\'ena, Abgrall,
  Bize, and Wolf}}]{Hees:2016gop}
\bibinfo{author}{\bibfnamefont{A.}~\bibnamefont{Hees}},
  \bibinfo{author}{\bibfnamefont{J.}~\bibnamefont{Gu\'ena}},
  \bibinfo{author}{\bibfnamefont{M.}~\bibnamefont{Abgrall}},
  \bibinfo{author}{\bibfnamefont{S.}~\bibnamefont{Bize}}, \bibnamefont{and}
  \bibinfo{author}{\bibfnamefont{P.}~\bibnamefont{Wolf}},
  \bibinfo{journal}{Phys. Rev. Lett.} \textbf{\bibinfo{volume}{117}},
  \bibinfo{pages}{061301} (\bibinfo{year}{2016}), \eprint{1604.08514}.

\bibitem[{\citenamefont{Kouvaris et~al.}(2020)\citenamefont{Kouvaris,
  Papantonopoulos, Street, and Wijewardhana}}]{Kouvaris:2019nzd}
\bibinfo{author}{\bibfnamefont{C.}~\bibnamefont{Kouvaris}},
  \bibinfo{author}{\bibfnamefont{E.}~\bibnamefont{Papantonopoulos}},
  \bibinfo{author}{\bibfnamefont{L.}~\bibnamefont{Street}}, \bibnamefont{and}
  \bibinfo{author}{\bibfnamefont{L.~C.~R.} \bibnamefont{Wijewardhana}},
  \bibinfo{journal}{Phys. Rev. D} \textbf{\bibinfo{volume}{102}},
  \bibinfo{pages}{063014} (\bibinfo{year}{2020}), \eprint{1910.00567}.

\bibitem[{\citenamefont{Filzinger et~al.}(2023)\citenamefont{Filzinger,
  D\"orscher, Lange, Klose, Steinel, Benkler, Peik, Lisdat, and
  Huntemann}}]{Filzinger:2023zrs}
\bibinfo{author}{\bibfnamefont{M.}~\bibnamefont{Filzinger}},
  \bibinfo{author}{\bibfnamefont{S.}~\bibnamefont{D\"orscher}},
  \bibinfo{author}{\bibfnamefont{R.}~\bibnamefont{Lange}},
  \bibinfo{author}{\bibfnamefont{J.}~\bibnamefont{Klose}},
  \bibinfo{author}{\bibfnamefont{M.}~\bibnamefont{Steinel}},
  \bibinfo{author}{\bibfnamefont{E.}~\bibnamefont{Benkler}},
  \bibinfo{author}{\bibfnamefont{E.}~\bibnamefont{Peik}},
  \bibinfo{author}{\bibfnamefont{C.}~\bibnamefont{Lisdat}}, \bibnamefont{and}
  \bibinfo{author}{\bibfnamefont{N.}~\bibnamefont{Huntemann}},
  \bibinfo{journal}{Phys. Rev. Lett.} \textbf{\bibinfo{volume}{130}},
  \bibinfo{pages}{253001} (\bibinfo{year}{2023}), \eprint{2301.03433}.

\bibitem[{\citenamefont{Aiello et~al.}(2022)\citenamefont{Aiello, Richardson,
  Vermeulen, Grote, Hogan, Kwon, and Stoughton}}]{Aiello:2021wlp}
\bibinfo{author}{\bibfnamefont{L.}~\bibnamefont{Aiello}},
  \bibinfo{author}{\bibfnamefont{J.~W.} \bibnamefont{Richardson}},
  \bibinfo{author}{\bibfnamefont{S.~M.} \bibnamefont{Vermeulen}},
  \bibinfo{author}{\bibfnamefont{H.}~\bibnamefont{Grote}},
  \bibinfo{author}{\bibfnamefont{C.}~\bibnamefont{Hogan}},
  \bibinfo{author}{\bibfnamefont{O.}~\bibnamefont{Kwon}}, \bibnamefont{and}
  \bibinfo{author}{\bibfnamefont{C.}~\bibnamefont{Stoughton}},
  \bibinfo{journal}{Phys. Rev. Lett.} \textbf{\bibinfo{volume}{128}},
  \bibinfo{pages}{121101} (\bibinfo{year}{2022}), \eprint{2108.04746}.

\bibitem[{\citenamefont{Zhao et~al.}(2022)\citenamefont{Zhao, Gao, Wang, and
  Zhan}}]{Zhao:2021tie}
\bibinfo{author}{\bibfnamefont{W.}~\bibnamefont{Zhao}},
  \bibinfo{author}{\bibfnamefont{D.}~\bibnamefont{Gao}},
  \bibinfo{author}{\bibfnamefont{J.}~\bibnamefont{Wang}}, \bibnamefont{and}
  \bibinfo{author}{\bibfnamefont{M.}~\bibnamefont{Zhan}},
  \bibinfo{journal}{Gen. Rel. Grav.} \textbf{\bibinfo{volume}{54}},
  \bibinfo{pages}{41} (\bibinfo{year}{2022}), \eprint{2102.02391}.

\bibitem[{\citenamefont{Kim}(2023)}]{Kim:2023pkx}
\bibinfo{author}{\bibfnamefont{H.}~\bibnamefont{Kim}} (\bibinfo{year}{2023}),
  \eprint{2306.13348}.

\bibitem[{\citenamefont{Badurina et~al.}(2023)\citenamefont{Badurina, Gibson,
  McCabe, and Mitchell}}]{Badurina:2022ngn}
\bibinfo{author}{\bibfnamefont{L.}~\bibnamefont{Badurina}},
  \bibinfo{author}{\bibfnamefont{V.}~\bibnamefont{Gibson}},
  \bibinfo{author}{\bibfnamefont{C.}~\bibnamefont{McCabe}}, \bibnamefont{and}
  \bibinfo{author}{\bibfnamefont{J.}~\bibnamefont{Mitchell}},
  \bibinfo{journal}{Phys. Rev. D} \textbf{\bibinfo{volume}{107}},
  \bibinfo{pages}{055002} (\bibinfo{year}{2023}), \eprint{2211.01854}.

\bibitem[{\citenamefont{Cordero et~al.}(2023)\citenamefont{Cordero, Delgadillo,
  and Miranda}}]{Cordero:2022fwb}
\bibinfo{author}{\bibfnamefont{R.}~\bibnamefont{Cordero}},
  \bibinfo{author}{\bibfnamefont{L.~A.} \bibnamefont{Delgadillo}},
  \bibnamefont{and} \bibinfo{author}{\bibfnamefont{O.~G.}
  \bibnamefont{Miranda}}, \bibinfo{journal}{Phys. Rev. D}
  \textbf{\bibinfo{volume}{107}}, \bibinfo{pages}{075023}
  (\bibinfo{year}{2023}), \eprint{2207.11308}.

\bibitem[{\citenamefont{D'Antonio et~al.}(2018)}]{DAntonio:2018sff}
\bibinfo{author}{\bibfnamefont{S.}~\bibnamefont{D'Antonio}}
  \bibnamefont{et~al.}, \bibinfo{journal}{Phys. Rev. D}
  \textbf{\bibinfo{volume}{98}}, \bibinfo{pages}{103017}
  (\bibinfo{year}{2018}), \eprint{1809.07202}.

\bibitem[{\citenamefont{Isi et~al.}(2019)\citenamefont{Isi, Sun, Brito, and
  Melatos}}]{Isi:2018pzk}
\bibinfo{author}{\bibfnamefont{M.}~\bibnamefont{Isi}},
  \bibinfo{author}{\bibfnamefont{L.}~\bibnamefont{Sun}},
  \bibinfo{author}{\bibfnamefont{R.}~\bibnamefont{Brito}}, \bibnamefont{and}
  \bibinfo{author}{\bibfnamefont{A.}~\bibnamefont{Melatos}},
  \bibinfo{journal}{Phys. Rev. D} \textbf{\bibinfo{volume}{99}},
  \bibinfo{pages}{084042} (\bibinfo{year}{2019}), \bibinfo{note}{[Erratum:
  Phys.Rev.D 102, 049901 (2020)]}, \eprint{1810.03812}.

\bibitem[{\citenamefont{Morisaki and Suyama}(2019)}]{Morisaki:2018htj}
\bibinfo{author}{\bibfnamefont{S.}~\bibnamefont{Morisaki}} \bibnamefont{and}
  \bibinfo{author}{\bibfnamefont{T.}~\bibnamefont{Suyama}},
  \bibinfo{journal}{Phys. Rev. D} \textbf{\bibinfo{volume}{100}},
  \bibinfo{pages}{123512} (\bibinfo{year}{2019}), \eprint{1811.05003}.

\bibitem[{\citenamefont{Palomba et~al.}(2019)}]{Palomba:2019vxe}
\bibinfo{author}{\bibfnamefont{C.}~\bibnamefont{Palomba}} \bibnamefont{et~al.},
  \bibinfo{journal}{Phys. Rev. Lett.} \textbf{\bibinfo{volume}{123}},
  \bibinfo{pages}{171101} (\bibinfo{year}{2019}), \eprint{1909.08854}.

\bibitem[{\citenamefont{Sun et~al.}(2020)\citenamefont{Sun, Brito, and
  Isi}}]{Sun:2019mqb}
\bibinfo{author}{\bibfnamefont{L.}~\bibnamefont{Sun}},
  \bibinfo{author}{\bibfnamefont{R.}~\bibnamefont{Brito}}, \bibnamefont{and}
  \bibinfo{author}{\bibfnamefont{M.}~\bibnamefont{Isi}},
  \bibinfo{journal}{Phys. Rev. D} \textbf{\bibinfo{volume}{101}},
  \bibinfo{pages}{063020} (\bibinfo{year}{2020}), \bibinfo{note}{[Erratum:
  Phys.Rev.D 102, 089902 (2020)]}, \eprint{1909.11267}.

\bibitem[{\citenamefont{Ng et~al.}(2020)\citenamefont{Ng, Isi, Haster, and
  Vitale}}]{Ng:2020jqd}
\bibinfo{author}{\bibfnamefont{K.~K.~Y.} \bibnamefont{Ng}},
  \bibinfo{author}{\bibfnamefont{M.}~\bibnamefont{Isi}},
  \bibinfo{author}{\bibfnamefont{C.-J.} \bibnamefont{Haster}},
  \bibnamefont{and} \bibinfo{author}{\bibfnamefont{S.}~\bibnamefont{Vitale}},
  \bibinfo{journal}{Phys. Rev. D} \textbf{\bibinfo{volume}{102}},
  \bibinfo{pages}{083020} (\bibinfo{year}{2020}), \eprint{2007.12793}.

\bibitem[{\citenamefont{Ng et~al.}(2021{\natexlab{a}})\citenamefont{Ng, Vitale,
  Hannuksela, and Li}}]{Ng:2020ruv}
\bibinfo{author}{\bibfnamefont{K.~K.~Y.} \bibnamefont{Ng}},
  \bibinfo{author}{\bibfnamefont{S.}~\bibnamefont{Vitale}},
  \bibinfo{author}{\bibfnamefont{O.~A.} \bibnamefont{Hannuksela}},
  \bibnamefont{and} \bibinfo{author}{\bibfnamefont{T.~G.~F.} \bibnamefont{Li}},
  \bibinfo{journal}{Phys. Rev. Lett.} \textbf{\bibinfo{volume}{126}},
  \bibinfo{pages}{151102} (\bibinfo{year}{2021}{\natexlab{a}}),
  \eprint{2011.06010}.

\bibitem[{\citenamefont{Banerjee et~al.}(2023)\citenamefont{Banerjee, Bera, and
  Mota}}]{Banerjee:2022zii}
\bibinfo{author}{\bibfnamefont{S.}~\bibnamefont{Banerjee}},
  \bibinfo{author}{\bibfnamefont{S.}~\bibnamefont{Bera}}, \bibnamefont{and}
  \bibinfo{author}{\bibfnamefont{D.~F.} \bibnamefont{Mota}},
  \bibinfo{journal}{JCAP} \textbf{\bibinfo{volume}{03}}, \bibinfo{pages}{041}
  (\bibinfo{year}{2023}), \eprint{2211.13988}.

\bibitem[{\citenamefont{Liu et~al.}(2021)\citenamefont{Liu, Yang, Wu, Xing, Xu,
  and Long}}]{Liu:2021xfb}
\bibinfo{author}{\bibfnamefont{D.}~\bibnamefont{Liu}},
  \bibinfo{author}{\bibfnamefont{Y.}~\bibnamefont{Yang}},
  \bibinfo{author}{\bibfnamefont{S.}~\bibnamefont{Wu}},
  \bibinfo{author}{\bibfnamefont{Y.}~\bibnamefont{Xing}},
  \bibinfo{author}{\bibfnamefont{Z.}~\bibnamefont{Xu}}, \bibnamefont{and}
  \bibinfo{author}{\bibfnamefont{Z.-W.} \bibnamefont{Long}},
  \bibinfo{journal}{Phys. Rev. D} \textbf{\bibinfo{volume}{104}},
  \bibinfo{pages}{104042} (\bibinfo{year}{2021}), \eprint{2104.04332}.

\bibitem[{\citenamefont{Manita et~al.}(2023)\citenamefont{Manita, Takeda, Aoki,
  Fujita, and Mukohyama}}]{Manita:2023mnc}
\bibinfo{author}{\bibfnamefont{Y.}~\bibnamefont{Manita}},
  \bibinfo{author}{\bibfnamefont{H.}~\bibnamefont{Takeda}},
  \bibinfo{author}{\bibfnamefont{K.}~\bibnamefont{Aoki}},
  \bibinfo{author}{\bibfnamefont{T.}~\bibnamefont{Fujita}}, \bibnamefont{and}
  \bibinfo{author}{\bibfnamefont{S.}~\bibnamefont{Mukohyama}}
  (\bibinfo{year}{2023}), \eprint{2310.10646}.

\bibitem[{\citenamefont{Yu et~al.}(2023)\citenamefont{Yu, Yao, Tang, and
  Wu}}]{Yu:2023iog}
\bibinfo{author}{\bibfnamefont{J.-C.} \bibnamefont{Yu}},
  \bibinfo{author}{\bibfnamefont{Y.-H.} \bibnamefont{Yao}},
  \bibinfo{author}{\bibfnamefont{Y.}~\bibnamefont{Tang}}, \bibnamefont{and}
  \bibinfo{author}{\bibfnamefont{Y.-L.} \bibnamefont{Wu}},
  \bibinfo{journal}{Phys. Rev. D} \textbf{\bibinfo{volume}{108}},
  \bibinfo{pages}{083007} (\bibinfo{year}{2023}), \eprint{2307.09197}.

\bibitem[{\citenamefont{Miller and Mendes}(2023)}]{Miller:2023kkd}
\bibinfo{author}{\bibfnamefont{A.~L.} \bibnamefont{Miller}} \bibnamefont{and}
  \bibinfo{author}{\bibfnamefont{L.}~\bibnamefont{Mendes}},
  \bibinfo{journal}{Phys. Rev. D} \textbf{\bibinfo{volume}{107}},
  \bibinfo{pages}{063015} (\bibinfo{year}{2023}), \eprint{2301.08736}.

\bibitem[{\citenamefont{Tsutsui and Nishizawa}(2023)}]{Tsutsui:2023jbk}
\bibinfo{author}{\bibfnamefont{T.}~\bibnamefont{Tsutsui}} \bibnamefont{and}
  \bibinfo{author}{\bibfnamefont{A.}~\bibnamefont{Nishizawa}},
  \bibinfo{journal}{Phys. Rev. D} \textbf{\bibinfo{volume}{107}},
  \bibinfo{pages}{103516} (\bibinfo{year}{2023}), \eprint{2303.07688}.

\bibitem[{\citenamefont{Delgado}(2023)}]{Delgado:2023psl}
\bibinfo{author}{\bibfnamefont{P.~C.~M.} \bibnamefont{Delgado}}
  (\bibinfo{year}{2023}), \eprint{2309.09946}.

\bibitem[{\citenamefont{Hannuksela et~al.}(2019)\citenamefont{Hannuksela, Wong,
  Brito, Berti, and Li}}]{Hannuksela:2018izj}
\bibinfo{author}{\bibfnamefont{O.~A.} \bibnamefont{Hannuksela}},
  \bibinfo{author}{\bibfnamefont{K.~W.~K.} \bibnamefont{Wong}},
  \bibinfo{author}{\bibfnamefont{R.}~\bibnamefont{Brito}},
  \bibinfo{author}{\bibfnamefont{E.}~\bibnamefont{Berti}}, \bibnamefont{and}
  \bibinfo{author}{\bibfnamefont{T.~G.~F.} \bibnamefont{Li}},
  \bibinfo{journal}{Nature Astron.} \textbf{\bibinfo{volume}{3}},
  \bibinfo{pages}{447} (\bibinfo{year}{2019}), \eprint{1804.09659}.

\bibitem[{\citenamefont{Ng et~al.}(2021{\natexlab{b}})\citenamefont{Ng,
  Hannuksela, Vitale, and Li}}]{Ng:2019jsx}
\bibinfo{author}{\bibfnamefont{K.~K.~Y.} \bibnamefont{Ng}},
  \bibinfo{author}{\bibfnamefont{O.~A.} \bibnamefont{Hannuksela}},
  \bibinfo{author}{\bibfnamefont{S.}~\bibnamefont{Vitale}}, \bibnamefont{and}
  \bibinfo{author}{\bibfnamefont{T.~G.~F.} \bibnamefont{Li}},
  \bibinfo{journal}{Phys. Rev. D} \textbf{\bibinfo{volume}{103}},
  \bibinfo{pages}{063010} (\bibinfo{year}{2021}{\natexlab{b}}),
  \eprint{1908.02312}.

\bibitem[{\citenamefont{Chung et~al.}(2021)\citenamefont{Chung, Gais, Cheung,
  and Li}}]{Chung:2021roh}
\bibinfo{author}{\bibfnamefont{A.~K.-W.} \bibnamefont{Chung}},
  \bibinfo{author}{\bibfnamefont{J.}~\bibnamefont{Gais}},
  \bibinfo{author}{\bibfnamefont{M.~H.-Y.} \bibnamefont{Cheung}},
  \bibnamefont{and} \bibinfo{author}{\bibfnamefont{T.~G.~F.} \bibnamefont{Li}},
  \bibinfo{journal}{Phys. Rev. D} \textbf{\bibinfo{volume}{104}},
  \bibinfo{pages}{084028} (\bibinfo{year}{2021}), \eprint{2107.05492}.

\bibitem[{\citenamefont{Chan and Hannuksela}(2022)}]{Chan:2022dkt}
\bibinfo{author}{\bibfnamefont{K.~H.~M.} \bibnamefont{Chan}} \bibnamefont{and}
  \bibinfo{author}{\bibfnamefont{O.~A.} \bibnamefont{Hannuksela}}
  (\bibinfo{year}{2022}), \eprint{2209.03536}.

\bibitem[{\citenamefont{Pantig and \"Ovg\"un}(2022)}]{Pantig:2022toh}
\bibinfo{author}{\bibfnamefont{R.~C.} \bibnamefont{Pantig}} \bibnamefont{and}
  \bibinfo{author}{\bibfnamefont{A.}~\bibnamefont{\"Ovg\"un}},
  \bibinfo{journal}{Eur. Phys. J. C} \textbf{\bibinfo{volume}{82}},
  \bibinfo{pages}{391} (\bibinfo{year}{2022}), \eprint{2201.03365}.

\bibitem[{\citenamefont{Qin et~al.}(2022)\citenamefont{Qin, Zhao, Du, Ke, Luo,
  Tan, and Shao}}]{Qin:2022qnk}
\bibinfo{author}{\bibfnamefont{C.}~\bibnamefont{Qin}},
  \bibinfo{author}{\bibfnamefont{B.}~\bibnamefont{Zhao}},
  \bibinfo{author}{\bibfnamefont{A.}~\bibnamefont{Du}},
  \bibinfo{author}{\bibfnamefont{J.}~\bibnamefont{Ke}},
  \bibinfo{author}{\bibfnamefont{J.}~\bibnamefont{Luo}},
  \bibinfo{author}{\bibfnamefont{Y.}~\bibnamefont{Tan}}, \bibnamefont{and}
  \bibinfo{author}{\bibfnamefont{C.}~\bibnamefont{Shao}}
  (\bibinfo{year}{2022}), \eprint{2212.06032}.

\bibitem[{\citenamefont{Traykova et~al.}(2021)\citenamefont{Traykova, Clough,
  Helfer, Berti, Ferreira, and Hui}}]{Traykova:2021dua}
\bibinfo{author}{\bibfnamefont{D.}~\bibnamefont{Traykova}},
  \bibinfo{author}{\bibfnamefont{K.}~\bibnamefont{Clough}},
  \bibinfo{author}{\bibfnamefont{T.}~\bibnamefont{Helfer}},
  \bibinfo{author}{\bibfnamefont{E.}~\bibnamefont{Berti}},
  \bibinfo{author}{\bibfnamefont{P.~G.} \bibnamefont{Ferreira}},
  \bibnamefont{and} \bibinfo{author}{\bibfnamefont{L.}~\bibnamefont{Hui}},
  \bibinfo{journal}{Phys. Rev. D} \textbf{\bibinfo{volume}{104}},
  \bibinfo{pages}{103014} (\bibinfo{year}{2021}), \eprint{2106.08280}.

\bibitem[{\citenamefont{Wang and Easther}(2022)}]{Wang:2021udl}
\bibinfo{author}{\bibfnamefont{Y.}~\bibnamefont{Wang}} \bibnamefont{and}
  \bibinfo{author}{\bibfnamefont{R.}~\bibnamefont{Easther}},
  \bibinfo{journal}{Phys. Rev. D} \textbf{\bibinfo{volume}{105}},
  \bibinfo{pages}{063523} (\bibinfo{year}{2022}), \eprint{2110.03428}.

\bibitem[{\citenamefont{Vicente and Cardoso}(2022)}]{Vicente:2022ivh}
\bibinfo{author}{\bibfnamefont{R.}~\bibnamefont{Vicente}} \bibnamefont{and}
  \bibinfo{author}{\bibfnamefont{V.}~\bibnamefont{Cardoso}},
  \bibinfo{journal}{Phys. Rev. D} \textbf{\bibinfo{volume}{105}},
  \bibinfo{pages}{083008} (\bibinfo{year}{2022}), \eprint{2201.08854}.

\bibitem[{\citenamefont{Boudon et~al.}(2022)\citenamefont{Boudon, Brax, and
  Valageas}}]{Boudon:2022dxi}
\bibinfo{author}{\bibfnamefont{A.}~\bibnamefont{Boudon}},
  \bibinfo{author}{\bibfnamefont{P.}~\bibnamefont{Brax}}, \bibnamefont{and}
  \bibinfo{author}{\bibfnamefont{P.}~\bibnamefont{Valageas}},
  \bibinfo{journal}{Phys. Rev. D} \textbf{\bibinfo{volume}{106}},
  \bibinfo{pages}{043507} (\bibinfo{year}{2022}), \eprint{2204.09401}.

\bibitem[{\citenamefont{Blas et~al.}(2020)\citenamefont{Blas, L\'opez~Nacir,
  and Sibiryakov}}]{Blas:2019hxz}
\bibinfo{author}{\bibfnamefont{D.}~\bibnamefont{Blas}},
  \bibinfo{author}{\bibfnamefont{D.}~\bibnamefont{L\'opez~Nacir}},
  \bibnamefont{and}
  \bibinfo{author}{\bibfnamefont{S.}~\bibnamefont{Sibiryakov}},
  \bibinfo{journal}{Phys. Rev. D} \textbf{\bibinfo{volume}{101}},
  \bibinfo{pages}{063016} (\bibinfo{year}{2020}), \eprint{1910.08544}.

\bibitem[{\citenamefont{Su et~al.}(2021)\citenamefont{Su, Xianyu, and
  Zhang}}]{Su:2021dwz}
\bibinfo{author}{\bibfnamefont{B.}~\bibnamefont{Su}},
  \bibinfo{author}{\bibfnamefont{Z.-Z.} \bibnamefont{Xianyu}},
  \bibnamefont{and} \bibinfo{author}{\bibfnamefont{X.}~\bibnamefont{Zhang}},
  \bibinfo{journal}{Astrophys. J.} \textbf{\bibinfo{volume}{923}},
  \bibinfo{pages}{114} (\bibinfo{year}{2021}), \eprint{2107.13527}.

\bibitem[{\citenamefont{Barsanti et~al.}(2023)\citenamefont{Barsanti, Maselli,
  Sotiriou, and Gualtieri}}]{Barsanti:2022vvl}
\bibinfo{author}{\bibfnamefont{S.}~\bibnamefont{Barsanti}},
  \bibinfo{author}{\bibfnamefont{A.}~\bibnamefont{Maselli}},
  \bibinfo{author}{\bibfnamefont{T.~P.} \bibnamefont{Sotiriou}},
  \bibnamefont{and}
  \bibinfo{author}{\bibfnamefont{L.}~\bibnamefont{Gualtieri}},
  \bibinfo{journal}{Phys. Rev. Lett.} \textbf{\bibinfo{volume}{131}},
  \bibinfo{pages}{051401} (\bibinfo{year}{2023}), \eprint{2212.03888}.

\bibitem[{\citenamefont{Davoudiasl and Denton}(2019)}]{Davoudiasl:2019nlo}
\bibinfo{author}{\bibfnamefont{H.}~\bibnamefont{Davoudiasl}} \bibnamefont{and}
  \bibinfo{author}{\bibfnamefont{P.~B.} \bibnamefont{Denton}},
  \bibinfo{journal}{Phys. Rev. Lett.} \textbf{\bibinfo{volume}{123}},
  \bibinfo{pages}{021102} (\bibinfo{year}{2019}), \eprint{1904.09242}.

\bibitem[{\citenamefont{Tsai et~al.}(2023{\natexlab{a}})\citenamefont{Tsai, Wu,
  Vagnozzi, and Visinelli}}]{Tsai:2021irw}
\bibinfo{author}{\bibfnamefont{Y.-D.} \bibnamefont{Tsai}},
  \bibinfo{author}{\bibfnamefont{Y.}~\bibnamefont{Wu}},
  \bibinfo{author}{\bibfnamefont{S.}~\bibnamefont{Vagnozzi}}, \bibnamefont{and}
  \bibinfo{author}{\bibfnamefont{L.}~\bibnamefont{Visinelli}},
  \bibinfo{journal}{JCAP} \textbf{\bibinfo{volume}{04}}, \bibinfo{pages}{031}
  (\bibinfo{year}{2023}{\natexlab{a}}), \eprint{2107.04038}.

\bibitem[{\citenamefont{Chakrabarti et~al.}(2022)\citenamefont{Chakrabarti,
  Dave, Dutta, and Goswami}}]{Chakrabarti:2022owq}
\bibinfo{author}{\bibfnamefont{S.}~\bibnamefont{Chakrabarti}},
  \bibinfo{author}{\bibfnamefont{B.}~\bibnamefont{Dave}},
  \bibinfo{author}{\bibfnamefont{K.}~\bibnamefont{Dutta}}, \bibnamefont{and}
  \bibinfo{author}{\bibfnamefont{G.}~\bibnamefont{Goswami}},
  \bibinfo{journal}{JCAP} \textbf{\bibinfo{volume}{09}}, \bibinfo{pages}{074}
  (\bibinfo{year}{2022}), \eprint{2202.11081}.

\bibitem[{\citenamefont{Tsai et~al.}(2023{\natexlab{b}})\citenamefont{Tsai,
  Farnocchia, Micheli, Vagnozzi, and Visinelli}}]{Tsai:2023zza}
\bibinfo{author}{\bibfnamefont{Y.-D.} \bibnamefont{Tsai}},
  \bibinfo{author}{\bibfnamefont{D.}~\bibnamefont{Farnocchia}},
  \bibinfo{author}{\bibfnamefont{M.}~\bibnamefont{Micheli}},
  \bibinfo{author}{\bibfnamefont{S.}~\bibnamefont{Vagnozzi}}, \bibnamefont{and}
  \bibinfo{author}{\bibfnamefont{L.}~\bibnamefont{Visinelli}}
  (\bibinfo{year}{2023}{\natexlab{b}}), \eprint{2309.13106}.

\bibitem[{\citenamefont{Laroche et~al.}(2022)\citenamefont{Laroche, Gilman, Li,
  Bovy, and Du}}]{Laroche:2022pjm}
\bibinfo{author}{\bibfnamefont{A.}~\bibnamefont{Laroche}},
  \bibinfo{author}{\bibfnamefont{D.}~\bibnamefont{Gilman}},
  \bibinfo{author}{\bibfnamefont{X.}~\bibnamefont{Li}},
  \bibinfo{author}{\bibfnamefont{J.}~\bibnamefont{Bovy}}, \bibnamefont{and}
  \bibinfo{author}{\bibfnamefont{X.}~\bibnamefont{Du}} (\bibinfo{year}{2022}),
  \eprint{2206.11269}.

\bibitem[{\citenamefont{Poddar}(2022)}]{Poddar:2021ose}
\bibinfo{author}{\bibfnamefont{T.~K.} \bibnamefont{Poddar}},
  \bibinfo{journal}{Eur. Phys. J. C} \textbf{\bibinfo{volume}{82}},
  \bibinfo{pages}{982} (\bibinfo{year}{2022}), \eprint{2111.05632}.

\bibitem[{\citenamefont{Farren et~al.}(2022)\citenamefont{Farren, Grin, Jaffe,
  Hlo\v{z}ek, and Marsh}}]{Farren:2021jcd}
\bibinfo{author}{\bibfnamefont{G.~S.} \bibnamefont{Farren}},
  \bibinfo{author}{\bibfnamefont{D.}~\bibnamefont{Grin}},
  \bibinfo{author}{\bibfnamefont{A.~H.} \bibnamefont{Jaffe}},
  \bibinfo{author}{\bibfnamefont{R.}~\bibnamefont{Hlo\v{z}ek}},
  \bibnamefont{and} \bibinfo{author}{\bibfnamefont{D.~J.~E.}
  \bibnamefont{Marsh}}, \bibinfo{journal}{Phys. Rev. D}
  \textbf{\bibinfo{volume}{105}}, \bibinfo{pages}{063513}
  (\bibinfo{year}{2022}), \eprint{2109.13268}.

\bibitem[{\citenamefont{Poddar}(2021)}]{Poddar:2021sbc}
\bibinfo{author}{\bibfnamefont{T.~K.} \bibnamefont{Poddar}},
  \bibinfo{journal}{JCAP} \textbf{\bibinfo{volume}{09}}, \bibinfo{pages}{041}
  (\bibinfo{year}{2021}), \eprint{2104.09772}.

\bibitem[{\citenamefont{Della~Monica and
  de~Martino}(2023)}]{DellaMonica:2023dcw}
\bibinfo{author}{\bibfnamefont{R.}~\bibnamefont{Della~Monica}}
  \bibnamefont{and}
  \bibinfo{author}{\bibfnamefont{I.}~\bibnamefont{de~Martino}},
  \bibinfo{journal}{Phys. Rev. D} \textbf{\bibinfo{volume}{108}},
  \bibinfo{pages}{L101303} (\bibinfo{year}{2023}), \eprint{2305.10242}.

\bibitem[{\citenamefont{Yuan et~al.}(2022)\citenamefont{Yuan, Shen, Tsai, Yuan,
  and Fan}}]{Yuan:2022nmu}
\bibinfo{author}{\bibfnamefont{G.-W.} \bibnamefont{Yuan}},
  \bibinfo{author}{\bibfnamefont{Z.-Q.} \bibnamefont{Shen}},
  \bibinfo{author}{\bibfnamefont{Y.-L.~S.} \bibnamefont{Tsai}},
  \bibinfo{author}{\bibfnamefont{Q.}~\bibnamefont{Yuan}}, \bibnamefont{and}
  \bibinfo{author}{\bibfnamefont{Y.-Z.} \bibnamefont{Fan}},
  \bibinfo{journal}{Phys. Rev. D} \textbf{\bibinfo{volume}{106}},
  \bibinfo{pages}{103024} (\bibinfo{year}{2022}), \eprint{2205.04970}.

\bibitem[{\citenamefont{Porayko et~al.}(2018)}]{Porayko:2018sfa}
\bibinfo{author}{\bibfnamefont{N.~K.} \bibnamefont{Porayko}}
  \bibnamefont{et~al.}, \bibinfo{journal}{Phys. Rev. D}
  \textbf{\bibinfo{volume}{98}}, \bibinfo{pages}{102002}
  (\bibinfo{year}{2018}), \eprint{1810.03227}.

\bibitem[{\citenamefont{Afzal et~al.}(2023)}]{NANOGrav:2023hvm}
\bibinfo{author}{\bibfnamefont{A.}~\bibnamefont{Afzal}} \bibnamefont{et~al.}
  (\bibinfo{collaboration}{NANOGrav}), \bibinfo{journal}{Astrophys. J. Lett.}
  \textbf{\bibinfo{volume}{951}}, \bibinfo{pages}{L11} (\bibinfo{year}{2023}),
  \eprint{2306.16219}.

\bibitem[{\citenamefont{Antoniadis et~al.}(2023)}]{EPTA:2023xxk}
\bibinfo{author}{\bibfnamefont{J.}~\bibnamefont{Antoniadis}}
  \bibnamefont{et~al.} (\bibinfo{collaboration}{EPTA}) (\bibinfo{year}{2023}),
  \eprint{2306.16227}.

\bibitem[{\citenamefont{Hwang et~al.}(2023)\citenamefont{Hwang, Jeong, Noh, and
  Smarra}}]{Hwang:2023odi}
\bibinfo{author}{\bibfnamefont{J.-c.} \bibnamefont{Hwang}},
  \bibinfo{author}{\bibfnamefont{D.}~\bibnamefont{Jeong}},
  \bibinfo{author}{\bibfnamefont{H.}~\bibnamefont{Noh}}, \bibnamefont{and}
  \bibinfo{author}{\bibfnamefont{C.}~\bibnamefont{Smarra}}
  (\bibinfo{year}{2023}), \eprint{2311.00234}.

\bibitem[{\citenamefont{Smarra
  et~al.}(2023)}]{EuropeanPulsarTimingArray:2023egv}
\bibinfo{author}{\bibfnamefont{C.}~\bibnamefont{Smarra}} \bibnamefont{et~al.}
  (\bibinfo{collaboration}{European Pulsar Timing Array}),
  \bibinfo{journal}{Phys. Rev. Lett.} \textbf{\bibinfo{volume}{131}},
  \bibinfo{pages}{171001} (\bibinfo{year}{2023}), \eprint{2306.16228}.

\bibitem[{\citenamefont{Xia et~al.}(2023)\citenamefont{Xia, Tang, Huang, Yuan,
  and Fan}}]{Xia:2023hov}
\bibinfo{author}{\bibfnamefont{Z.-Q.} \bibnamefont{Xia}},
  \bibinfo{author}{\bibfnamefont{T.-P.} \bibnamefont{Tang}},
  \bibinfo{author}{\bibfnamefont{X.}~\bibnamefont{Huang}},
  \bibinfo{author}{\bibfnamefont{Q.}~\bibnamefont{Yuan}}, \bibnamefont{and}
  \bibinfo{author}{\bibfnamefont{Y.-Z.} \bibnamefont{Fan}},
  \bibinfo{journal}{Phys. Rev. D} \textbf{\bibinfo{volume}{107}},
  \bibinfo{pages}{L121302} (\bibinfo{year}{2023}), \eprint{2303.17545}.

\bibitem[{\citenamefont{De~Martino et~al.}(2020)\citenamefont{De~Martino,
  Broadhurst, Tye, Chiueh, and Schive}}]{DeMartino:2018zkx}
\bibinfo{author}{\bibfnamefont{I.}~\bibnamefont{De~Martino}},
  \bibinfo{author}{\bibfnamefont{T.}~\bibnamefont{Broadhurst}},
  \bibinfo{author}{\bibfnamefont{S.~H.~H.} \bibnamefont{Tye}},
  \bibinfo{author}{\bibfnamefont{T.}~\bibnamefont{Chiueh}}, \bibnamefont{and}
  \bibinfo{author}{\bibfnamefont{H.-Y.} \bibnamefont{Schive}},
  \bibinfo{journal}{Phys. Dark Univ.} \textbf{\bibinfo{volume}{28}},
  \bibinfo{pages}{100503} (\bibinfo{year}{2020}), \eprint{1807.08153}.

\bibitem[{\citenamefont{Dave and Goswami}(2023)}]{Dave:2023egr}
\bibinfo{author}{\bibfnamefont{B.}~\bibnamefont{Dave}} \bibnamefont{and}
  \bibinfo{author}{\bibfnamefont{G.}~\bibnamefont{Goswami}}
  (\bibinfo{year}{2023}), \eprint{2310.19664}.

\bibitem[{\citenamefont{Ir\v{s}i\v{c} et~al.}(2017)\citenamefont{Ir\v{s}i\v{c},
  Viel, Haehnelt, Bolton, and Becker}}]{Irsic:2017yje}
\bibinfo{author}{\bibfnamefont{V.}~\bibnamefont{Ir\v{s}i\v{c}}},
  \bibinfo{author}{\bibfnamefont{M.}~\bibnamefont{Viel}},
  \bibinfo{author}{\bibfnamefont{M.~G.} \bibnamefont{Haehnelt}},
  \bibinfo{author}{\bibfnamefont{J.~S.} \bibnamefont{Bolton}},
  \bibnamefont{and} \bibinfo{author}{\bibfnamefont{G.~D.}
  \bibnamefont{Becker}}, \bibinfo{journal}{Phys. Rev. Lett.}
  \textbf{\bibinfo{volume}{119}}, \bibinfo{pages}{031302}
  (\bibinfo{year}{2017}), \eprint{1703.04683}.

\bibitem[{\citenamefont{Armengaud et~al.}(2017)\citenamefont{Armengaud,
  Palanque-Delabrouille, Y\`eche, Marsh, and Baur}}]{Armengaud:2017nkf}
\bibinfo{author}{\bibfnamefont{E.}~\bibnamefont{Armengaud}},
  \bibinfo{author}{\bibfnamefont{N.}~\bibnamefont{Palanque-Delabrouille}},
  \bibinfo{author}{\bibfnamefont{C.}~\bibnamefont{Y\`eche}},
  \bibinfo{author}{\bibfnamefont{D.~J.~E.} \bibnamefont{Marsh}},
  \bibnamefont{and} \bibinfo{author}{\bibfnamefont{J.}~\bibnamefont{Baur}},
  \bibinfo{journal}{Mon. Not. Roy. Astron. Soc.}
  \textbf{\bibinfo{volume}{471}}, \bibinfo{pages}{4606} (\bibinfo{year}{2017}),
  \eprint{1703.09126}.

\bibitem[{\citenamefont{Nori et~al.}(2019)\citenamefont{Nori, Murgia,
  Ir\v{s}i\v{c}, Baldi, and Viel}}]{Nori:2018pka}
\bibinfo{author}{\bibfnamefont{M.}~\bibnamefont{Nori}},
  \bibinfo{author}{\bibfnamefont{R.}~\bibnamefont{Murgia}},
  \bibinfo{author}{\bibfnamefont{V.}~\bibnamefont{Ir\v{s}i\v{c}}},
  \bibinfo{author}{\bibfnamefont{M.}~\bibnamefont{Baldi}}, \bibnamefont{and}
  \bibinfo{author}{\bibfnamefont{M.}~\bibnamefont{Viel}},
  \bibinfo{journal}{Mon. Not. Roy. Astron. Soc.}
  \textbf{\bibinfo{volume}{482}}, \bibinfo{pages}{3227} (\bibinfo{year}{2019}),
  \eprint{1809.09619}.

\bibitem[{\citenamefont{Schive and Chiueh}(2018)}]{Schive:2017biq}
\bibinfo{author}{\bibfnamefont{H.-Y.} \bibnamefont{Schive}} \bibnamefont{and}
  \bibinfo{author}{\bibfnamefont{T.}~\bibnamefont{Chiueh}},
  \bibinfo{journal}{Mon. Not. Roy. Astron. Soc.}
  \textbf{\bibinfo{volume}{473}}, \bibinfo{pages}{L36} (\bibinfo{year}{2018}),
  \eprint{1706.03723}.

\bibitem[{\citenamefont{Robles and Matos}(2013{\natexlab{b}})}]{Robles:2012kt}
\bibinfo{author}{\bibfnamefont{V.~H.} \bibnamefont{Robles}} \bibnamefont{and}
  \bibinfo{author}{\bibfnamefont{T.}~\bibnamefont{Matos}},
  \bibinfo{journal}{Astrophys. J.} \textbf{\bibinfo{volume}{763}},
  \bibinfo{pages}{19} (\bibinfo{year}{2013}{\natexlab{b}}), \eprint{1207.5858}.

\end{thebibliography}

\end{document}